\date{\today}

\documentclass[aps,prl,twocolumn,superscriptaddress,groupedaddress]{revtex4-2}  % for review and submission
\usepackage{dcolumn,graphicx,amsmath,amssymb,amsfonts,algorithm,algpseudocode}
\usepackage{qcircuit}
\usepackage{verbatim}
\usepackage[colorlinks=true,citecolor=magenta,urlcolor=magenta,linkcolor=magenta]{hyperref}

\newcommand{\pdiff}[2]{\frac{\partial #1}{\partial #2}}

\newcommand{\nqfd}[2]{n_{\mathrm{QFD}}}
\newcommand{\device}[1]{$\mathsf{ibmq}\_\mathsf{#1}$}
\newcommand{\nlayers}{n_{\mathrm{DF}}}
\newcommand{\nsteps}{n_{\mathrm{QFD}}}
\newcommand{\necho}{n_{\mathrm{echo}}}
\newcommand{\cnot}{$\mathsf{CNOT}$\,}
\newcommand{\numberop}[1]{\hat{n}_{#1}}
\newcommand{\zetaop}[1]{\hat{z}_{#1}}
\newcommand{\changeofbasis}[2]{\hat{G}_{#2 \leftarrow #1}}

\begin{document}

\title{
Quantum Filter Diagonalization with Double-Factorized Hamiltonians
}

\author{Jeffrey Cohn}
\email{jeffrey.cohn@ibm.com} 
\affiliation{
IBM Quantum, IBM Research – Almaden,  San Jose, CA 95120, USA
}

\author{Mario Motta}
\email{mario.motta@ibm.com} 
\affiliation{
IBM Quantum, IBM Research – Almaden,  San Jose, CA 95120, USA
}

\author{Robert M. Parrish}
\email{rob.parrish@qcware.com}
\affiliation{
QC Ware Corporation, Palo Alto, CA 94301, USA
}

\begin{abstract} 
We demonstrate a method that merges the quantum filter diagonalization (QFD) approach for hybrid quantum/classical solution of the time-independent electronic  Schr\"odinger equation with a low-rank double factorization (DF) approach for the representation of the electronic Hamiltonian. In particular, we explore the use of sparse ``compressed'' double factorization (C-DF) truncation of the Hamiltonian within the time-propagation elements of QFD, while retaining a similarly compressed but numerically converged double-factorized representation of the Hamiltonian for the operator expectation values needed in the QFD quantum matrix elements.
Together with significant circuit reduction optimizations and number-preserving post-selection/echo-sequencing error mitigation strategies, the method is found to provide accurate predictions for low-lying eigenspectra in a number of representative molecular systems, while requiring reasonably short circuit depths and modest measurement costs. The method is demonstrated by experiments on noise-free simulators, decoherence- and shot-noise including simulators, and real quantum hardware.
\end{abstract}

\maketitle

\section{Introduction}

Solving the many-particle Schr\"{o}dinger equation
to compute eigenpairs of a Hamiltonian operator is
an important application in computational science.
For example, it arises in the simulation of the
electronic structure of molecules and materials, 
as well as in mathematical optimization problems. %\cite{helgaker2012recent,farhi2001quantum}. 
In the context of classical computation, 
different strategies are employed to numerically determine 
approximate ground and excited Hamiltonian eigenpairs, 
typically by assuming that eigenstates have a certain structure.

Digital quantum computers have been proposed as 
an alternative and complementary approach to the
determination of approximate ground and excited Hamiltonian eigenstates.
While efficient ground- and excited-state 
determination cannot be guaranteed for a generic Hamiltonian, since this is a QMA-complete problem \cite{kempe2006complexity}, 
a wealth of heuristic quantum algorithms have been designed and demonstrated in recent years.
Simulation of the time-dependent Schr\"{o}dinger
equation, on the other hand, is a more natural 
application for a quantum computer, as it lies 
in the BQP complexity class \cite{georgescu2014quantum}.
This observation has generated an increasingly
intense research effort, aimed at integrating
the simulation of the time-dependent Schr\"{o}dinger
equation in the structure of quantum computational algorithms for eigenstate determination \cite{parrish2019quantum,stair2020multireference,klymko2021real}.

Quantum filter diagonalization (QFD) (and two similar methods developed simultaneously in the literature)  \cite{parrish2019quantum,huggins2020non,stair2020multireference} is a quantum
algorithm in which a Hamiltonian operator is 
projected on a subspace spanned by a set of 
non-orthogonal quantum states generated via 
approximate quantum time evolution (or other quantum circuit propagation), and post-facto classically diagonalized.
QFD can be regarded to as a quantum computational
equivalent of classical filter diagonalization \cite{neuhauser1990bound,neuhauser1994circumventing},
from which it inherits the connection with the
Lanczos algorithm. Furthermore, it is an example
of a quantum subspace diagonalization method \cite{mcclean2017hybrid,huggins2020non,motta2020determining,ollitrault2020quantum} in 
which, starting from a set of approximate reference 
states for the targeted eigenvectors that can be
easily prepared classically, a basis for a 
subspace is constructed using time propagation.

\begin{figure}[b]
\includegraphics[width=0.6\columnwidth]{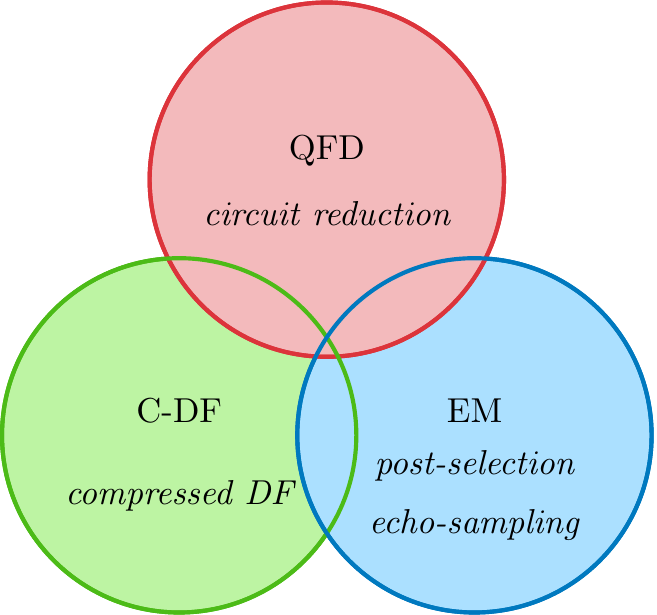}
\caption{
Schematic representation of the techniques explored in the present work: quantum filter diagonalization (QFD) is combined with a compressed double factorization (C-DF) approximation of the electronic Hamiltonian, and a set of error mitigation (EM) techniques.}
\label{fig:schema}
\end{figure}

The implementation of QFD on contemporary quantum
hardware poses a number of conceptual and technical
challenges. Among them, 
($i$) the high gate complexity required by the simulation 
of time evolution under the Hamiltonian, which is especially pronounced for the 
electronic structure problem, 
($ii$) the need of evaluating off-diagonal matrix elements 
that define overlap and Hamiltonian matrices, 
which requires Hadamard or swap quantum circuits, and 
($iii$) the integration of dedicated 
error mitigation techniques in the structure of 
the algorithm.

In the present work, we demonstrate a technique 
that merges QFD with a compressed low-rank double factorization 
of the electronic structure Hamiltonian, to achieve
substantially shorted circuit representations of the
time propagation steps, and to economize the evaluation
of off-diagonal matrix elements.
Further, we introduce a combination of post-selection 
and echo-sequencing aimed at mitigating errors from
violations of particle number and spin-$z$ conservation. 
The proposed techniques, sketched in Figure \ref{fig:schema}, 
are demonstrated using classical simulators of quantum devices, 
and performing experiments on IBM quantum hardware.

\section{Methods}

\subsection{``Compressed'' Double Factorized Electronic Hamiltonian}

A key technical element of this manuscript is the ``compressed'' double
factorization (C-DF) approximate representation of the electronic Hamiltonian,
which provides for reduced gate count requirements in quantum circuits for QFD
time propagation and reduced measurement requirements for Hamiltonian
expectation values. Below, we review the representation of the electronic Hamiltonian in the double-factorized representation as previously discussed by several authors in the literature \cite{
poulin2014trotter,
peng2017highly,
motta2018low,
motta2019efficient,
kivlichan2018quantum,
berry2019qubitization,
matsuzawa2020jastrow,
huggins2021efficient}, 
including the popular ``explicit'' double factorization procedure \cite{
motta2018low,
kivlichan2018quantum,
berry2019qubitization,
huggins2021efficient} (X-DF) for numerically finding the tensor factors. We then develop a new ``compressed'' double factorization (C-DF) procedure for numerically finding the tensor factors with enhanced compression and accuracy.

\subsubsection{The Electronic Hamiltonian}

In a real, orthonormal, and spin-restricted orbital basis (spatial parts of each
orbital $\{ \phi_p (\vec r_1) \}$ the same for each spin orbital $\{
\psi_{p,\alpha} (\vec x_1) \equiv \phi_{p} (\vec r_1) \alpha (s_1) \}$ and $\{
\phi_{p,\beta} (\vec x_1) \equiv \phi_{p} (\vec r_1) \beta (s_1) \}$), the
electronic Hamiltonian may be written as \cite{helgaker2014molecular}
\begin{equation}
\hat H
\equiv
E_{\mathrm{Ext}}
+
\sum_{pq}
(p|\hat \kappa|q)
\hat E_{pq}^{+}
+
\frac{1}{2}
\sum_{pqrs}
(pq|rs)
\hat E_{pq}^{+}
\hat E_{rs}^{+}
\end{equation}
where the singlet spin-summed 1-particle substitution operator is $\hat
E_{pq}^{+} \equiv \sum_{pq,\sigma} \hat p_{\sigma}^{\dagger} \hat q_{\sigma}$.

The spatial orbital electron repulsion integral (ERI) tensor is written in
chemists' notation as
\begin{equation}
(pq|rs)
\equiv
\iint_{\mathbb{R}^{6}}
\phi_{p} (\vec r_1)
\phi_{q} (\vec r_1)
\frac{1}{r_{12}}
\phi_{r} (\vec r_2)
\phi_{s} (\vec r_2)
.
\end{equation}
The modified spatial-orbital one-particle integrals are
\begin{equation}
(p|\hat \kappa|q)
\equiv
(p|\hat h|q)
-
\frac{1}{2}
\sum_{r}
(pr|qr)
.
\end{equation}
The spatial-orbital one-particle Hamiltonian integrals are
\begin{equation}
(p|\hat h|q)
\equiv
(p|-\nabla_{1}^{2}|q)
+
(p|-\sum_{A} Z_{A} / r_{1A}|q)
\end{equation}
\[
+
2 
\sum_{i} (pq|ii)
-
\sum_{i} (pi|qi)
.
\]
where $\{ \vec r_A \}$ are the nuclear positions, $\{ Z_A \}$ are the corresponding nuclear charges and $\{
\phi_{i} (\vec r_1) \}$ are the core spatial orbitals in an active space
picture. The external system self-energy is
\begin{equation}
E_{\mathrm{Ext}}
\equiv
\sum_{A > B}
\frac{Z_{A} Z_{B}}{r_{AB}}
.
\end{equation}
The external system self-energy, one-particle Hamiltonian integrals, and
electron repulsion integrals are polynomially tractable input quantities from
classical electronic structure codes.

When na\"ively expanded into Pauli words in e.g., the Jordan-Wigner
representation, this operator appears to require $\mathcal{O}(n_p^4)$ unique Pauli
words that occupy $\mathcal{O}(n_p^3)$ commuting groups, with significant efforts
required to minimize the prefactor of the $\mathcal{O}(n_p^3)$ group
determination \cite{
verteletskyi2020measurement,
gokhale2019minimizing}. This scaling implies significant computational cost in
time-propagation or evaluation of the expectation value of this Hamiltonian.
Recently, a number of authors have achieved substantial practical reductions in
this cost of \emph{both} of these considerations through an approach we will
generically refer to as ``double factorization'' (DF) \cite{
poulin2014trotter,
motta2018low,
kivlichan2018quantum,
berry2019qubitization,
matsuzawa2020jastrow,
huggins2021efficient}.\footnote{Note that this notation of double factorization $\rightarrow$ DF presents an unfortunate elision with the popular and related density fitting representation, which is often denoted as DF in the literature. To avoid any issues, we will explicitly write out ``density fitting'' for the few times it is encountered in this manuscript.}  The crux of the idea is a
representation of the ERI tensor as
\begin{equation}
\label{eq:DF}
(pq|rs)
\approx
\sum_{t}
\sum_{k}
\sum_{l}
U_{pk}^{t}
U_{qk}^{t}
Z_{kl}^{t}
U_{rl}^{t}
U_{sl}^{t}
.
\end{equation}
Where the ``leaf tensor'' $U_{pk}^{t}$ is constrained to be orthonormal (actually, we require this matrix to be special orthogonal without loss of generality for all cases encountered in this work)
\begin{equation}
\sum_{k}
U_{pk}^{t} U_{qk}^{t} = \delta_{pq}
,\
\sum_{k}
U_{kp}^{t} U_{kq}^{t} = \delta_{pq}
\ \forall \ t
,
\end{equation}
and the ``core tensor'' $Z_{pq}^{t}$ is constrained to be symmetric
\begin{equation}
Z_{kl}^{t}
=
Z_{lk}^{t}
\ \forall \ t
.
\end{equation}
In Equation \ref{eq:DF}, the sum over $t$ goes up to a maximum value of $n_{\mathrm{DF}}$, whose value is determined by the double factorization fitting procedure and user-defined input parameters.
Similarly, we can take the orthonormal eigendecomposition of
\begin{equation}
(p|\hat \kappa|q)
=
\sum_{k}
U_{pk}^{0}
f_{k}^{0}
U_{qk}^{0}
.
\end{equation}
Here only a single expansion index is needed due to the analytical nature of the eigendecomposition (in contrast to $n_{\mathrm{DF}}$ expansion indices for the ERI tensor above) and is denoted by the special index ``$^{0}$.''
Provided that efficient methods exist \cite{reck1994experimental,
wecker2015solving,
google2020hartree}
to transform the representation of the
one-particle orbital basis according to the orbital transformation $U_{pk}^{t}$,
the Hamiltonian may now be written as
\begin{equation}
\label{eq:df_hamiltonian}
\hat H
\equiv
E_{\mathrm{Ext}}
+
\sum_{k}
f_{k}^{0}
\hat E_{kk}^{+} (0)
+
\frac{1}{2}
\sum_{t}
\sum_{kl}
Z_{kl}^{t}
\hat E_{kk}^{+} (t)
\hat E_{ll}^{+} (t)
.
\end{equation}
The operators $\hat E_{kk}^{+} (t)$ are diagonal in qubit representations such as
the Jordan-Wigner and parity representations. Therefore, the expectations over
separate $k$ and $l$ terms can be measured simultaneously for each $t$,
drastically reducing the number of required measurements to evaluate Hamiltonian
expectation values. Similarly, the implementation of time-propagation
(Trotterized across $t$) can be accomplished by commuting and 
highly-parallelized controlled $\hat Z$ rotations. The generic idea of double
factorization can therefore substantially reduce the cost of quantum circuit
implementation of quantum chemistry methods. However, the specific practical
cost of implementation will depend on the number of $t$ factors required to
achieve an accurate factorization of the ERI tensor, i.e., on $n_{\mathrm{DF}}$. We try to reduce this cost
in the methodology developed below.

\subsubsection{``Explicit'' Double Factorization}

In the straightforward ``explicit'' approach to double factorization (which we label as X-DF) \cite{
motta2018low,
kivlichan2018quantum,
berry2019qubitization,
huggins2021efficient}, we first eigendecompose decompose the ERI tensor into a form of density fitting factorization:
\begin{equation}
(pq|rs)
=
\sum_{t}
V_{pq}^{t}
\lambda_{t}
V_{rs}^{t}
\end{equation}
where $V_{rs}^{t} = V_{sr}^{t}$ (up to numerical noise in accidental
degeneracies). Next, for each eigenvector, we eigendecompose
\begin{equation}
V_{rs}^{t}
=
\sum_{k}
U_{pk}^{t}
U_{qk}^{t}
\gamma_{k}^{t}
,
\end{equation}
and then form
\begin{equation}
Z_{kl}^{t}
\equiv
\gamma_{k}^{t}
\lambda_{t}
\gamma_{l}^{t}
.
\end{equation}

\subsubsection{``Compressed'' Double Factorization}

The X-DF procedure provides a straighforward and explicit recipe for obtaining the double-factorized quantities $U_{pk}^{t}$ and $Z_{kl}^{t}$, but it may be a non-optimal factorization due to the nested form of the eigendecompositions. Here, we consider an alternative ``compressed'' double factorization (C-DF) based on global optimization of a least squares objective function. 

The least-squares objective function for C-DF is
\begin{equation}
\mathcal{O} (U_{pk}^{t}, Z_{kl}^{t})
\end{equation}
\[
\equiv
\frac{1}{2}
\left |
(pq|rs)
-
\sum_{t}
\sum_{kl}
U_{pk}^{t}
U_{qk}^{t}
Z_{kl}^{t}
U_{rl}^{t}
U_{sl}^{t}
\right |_{\mathcal{F}}^2
.
\]
where $\mathcal{F}$ denotes the vector-type Frobenius norm.
Defining
\begin{equation}
\Delta_{pqrs}
\equiv
(pq|rs)
-
\sum_{t}
\sum_{kl}
U_{pk}^{t}
U_{qk}^{t}
Z_{kl}^{t}
U_{rl}^{t}
U_{sl}^{t}
,
\end{equation}
then the gradients are
\begin{equation}
\pdiff{\mathcal{O}}{Z_{kl}^{t}}
=
-
\sum_{pqrs}
\Delta_{pqrs}
U_{pk}^{t}
U_{qk}^{t}
U_{rl}^{t}
U_{sl}^{t}
\end{equation}
and
\begin{equation}
\pdiff{\mathcal{O}}{U_{pk}^{t}}
=
-
4
\sum_{qrsl}
\Delta_{pqrs}
U_{qk}^{t}
Z_{kl}^{t}
U_{rl}^{t}
U_{sl}^{t}
.
\end{equation}

Note that the a recent paper on Jastrow-Factor VQE \cite{matsuzawa2020jastrow} briefly considers in an appendix direct fitting of the doubly-factorized tensors of the ERI tensor to the exact ERI tensor, which goes beyond X-DF in the direction of C-DF. However, no details are given as to the numerical procedure used to perform this fit.

\textit{Unconstrained Form in Terms of Orbital Rotation Generators:} To remove the orthogonality constraints, one can always define the special orthogonal orbital rotation matrices $U_{pk}^{t}$ in terms of matrix exponentials of antisymmetric orbital rotation generator matrices $X_{pk}^{t}$
\begin{equation}
\label{eq:expm}
U_{pk}^{t}
\equiv
\left [
\exp(\hat X^t)
\right ]_{pk}
\end{equation}
where
\begin{equation}
\hat X^t
\equiv
X_{pq}^t
| p \rangle
\langle q |
\end{equation}
subject to
\begin{equation}
X_{pq}^t = - X_{qp}^t
\ \forall \ t
.
\end{equation}
In Equation \ref{eq:expm}, the notation $\exp(\hat M)$ means the matrix exponential of the matrix operator symbolically defined as $\hat M$.
In this form, the C-DF objective function becomes unconstrained
\begin{equation}
\mathcal{O} (U_{pk}^{t}, Z_{kl}^{t})
\rightarrow
\mathcal{O} (X_{pq}^{t}, Z_{kl}^{t})
,
\end{equation}
and the gradient of the C-DF objective function is easily evaluated through the chain rule, yielding
\begin{equation}
\pdiff{\mathcal{O}}{X_{k'l'}^{t}}
=
\sum_{pk}
\pdiff{\mathcal{O}}{U_{pk}^{t}}
\pdiff{U_{pk}^{t}}{X_{k'l'}^{t}}
.
\end{equation}
Efficient linear algebraic operations for the matrix exponential and the matrix
exponential gradient exist in the form of the Wilcox identity \cite{wilcox1967exponential}. These have been
specialized to the cases of antisymmetric generators $\hat
X$, and may be considered to be universal library functions for any $X_{pk}^{t}$. With this unconstrained formulation, one may supply the objective function and analytical gradient function to a numerical unconstrained continuous optimizer such as L-BFGS, and numerically optimize the $X_{pk}^{t}$ and $Z_{kl}^{t}$ factors of C-DF simultaneously. We have implemented this and found that while it provides a straightforward and simple approach, convergence can be markedly slow. For this reason, we pursue a nested ``two-step'' C-DF fitting procedure below.

\textit{Core Tensor Analytical Fitting:} An interesting avenue to explore is the form of the fitting equations when the
factors $U_{pk}^{t}$ are known (in analogy to the least-squares tensor hypercontraction procedure \cite{parrish2012tensor} in
non-orthogonal tensor hypercontraction, in which an analytical formula for $Z_{kl}$ resulted)
\begin{equation}
\mathcal{O}
(Z_{kl}^{t} | U_{pk}^{t})
=
\frac{1}{2}
\left |
(pq|rs)
-
\sum_{t}
\sum_{kl}
U_{pk}^{t}
U_{qk}^{t}
Z_{kl}^{t}
U_{rl}^{t}
U_{sl}^{t}
\right |_{\mathcal{F}}^2
.
\end{equation}
Here the weak form of the objective function is
\begin{equation}
\pdiff{\mathcal{O}}{Z_{kl}^{t}}
=
-
\sum_{pqrs}
\Delta_{pqrs}
U_{pk}^{t}
U_{qk}^{t}
U_{rl}^{t}
U_{sl}^{t}
=
0
\
\forall 
\ 
k, l, t
.
\end{equation}
Expanding yields
\begin{equation}
\sum_{t'}
\sum_{k'l'}
M_{kk'}^{tt'}
Z_{k'l'}^{t'}
M_{ll'}^{tt'}
=
R_{kl}^{t}
\
\forall 
\ 
k, l, t
\end{equation}
where
\begin{equation}
R_{kl}^{t}
\equiv
\sum_{pqrs}
U_{pk}^{t}
U_{qk}^{t}
(pq|rs)
U_{rl}^{t}
U_{sl}^{t}
\end{equation}
and
\begin{equation}
M_{kk'}^{tt'}
\equiv
\left [
\sum_{p}
U_{p k}^{t}
U_{p k'}^{t'}
\right ]
\left [
\sum_{q}
U_{q k}^{t}
U_{q k'}^{t'}
\right ]
\end{equation}
\[
=
\left [
\sum_{p}
U_{p k}^{t}
U_{p k'}^{t'}
\right ]^2
\equiv
\left [
S_{kk'}^{tt'}
\right ]^2
.
\]
The $^2$ notation is the element-wise square in the last expression. $\hat
S_{kk'}^{tt'}$ are metric matrices (symmetric, positive definite, with singular
values in $[0, 1]$) when unrolled in $kt \times k't'$. $M_{kk'}^{tt'}$ are thus
also metric matrices, with the extra specialization of having wholly positive
values.

These equations can be effectively written as
\begin{equation}
\label{eq:Zlinear}
\sum_{t'k'l'}
A_{tkl,t'k'l'}
Z_{t'k'l'}
=
R_{tkl}
\
\forall 
\ 
k, l, t
\end{equation}
where,
\begin{equation}
A_{tkl,t'k'l'}
\equiv
M_{kk'}^{tt'}
M_{ll'}^{tt'}
,
\end{equation}
i.e., a simple set of linear equations. It should be noted that the matrix $\hat
A$ has formal singularities of degeneracy $n_{\mathrm{DF}} - 1$, and may also contain
numerical near-singularities if the active $U_{pk}^{t}$ matrices are numerically
similar. In practice the eigendecomposition-based  Moore-Penrose pseudoinverse
approach to solve these equations costs $\mathcal{O}((n_{\mathrm{DF}} n_{p}^2)^3)$, which
is tractable for medium-sized problems. Moreover, conjugate gradient and L-BFGS
approaches appear to also provide reliable convergence with lowered cost. The
matrix-vector product primitive needed for such iterative approaches is
\begin{equation}
\sigma_{tkl}
\equiv
\sum_{t'k'l'}
A_{tkl,t'k'l'}
b_{t'k'l'}
.
\end{equation}
This can be efficiently implemented in terms of matrix multiplications with a
cost of $\mathcal{O} (n_{\mathrm{DF}}^2 n_{p}^{3})$.

\textit{Two-Step C-DF Fitting:} The above finding of an analytical fit for $Z_{kl}^{t}$ for any proposed
$U_{pk}^{t}$ leads to the following ``two-step'' C-DF fitting
\begin{equation}
\mathcal{O}
(U_{pk}^{t})
=
\frac{1}{2}
\left |
(pq|rs)
-
\sum_{t}
\sum_{kl}
U_{pk}^{t}
U_{qk}^{t}
Z_{kl}^{\downarrow t}
U_{rl}^{t}
U_{sl}^{t}
\right |_{\mathcal{F}}^2
.
\end{equation}
Where $Z_{kl}^{\downarrow t}$ is shorthand for the optimal $Z_{kl}^{t}$
predicated on the current $U_{pk}^{t}$ discussed in the section above.  In
practice, one actually works with the unconstrained form $\mathcal{O}
(X_{pq}^{t})$ within this two-step C-DF fitting procedure, with the
unconstrained optimization in $X_{pq}^{t}$ being handled by L-BFGS. The explicit two-step C-DF procedure is,

\textbf{Stage 0:} Use the X-DF factorization to obtain a guess for the factors $\{ U_{pk}^{t} \}$ and $\{ Z_{kl}^{t} \}$.

\textbf{Stage 1 (Optional - Will be first iteration of Stage 2):} For $\{
U_{pk}^{t} \}$ from Stage 1, find the globally optimal $\{ Z_{kl}^{t} \}$ via
least-squares (analytical).

\textbf{Stage 2:} For $U_{pk}^{t}$ from Stage 0 or Stage 1 (identical), run
two-step C-DF fitting to find the globally optimal $\{ U_{pk}^{t} \}$ and $\{
Z_{kl}^{t} \}$.

This procedure is analogous to the ``two-step CASSCF'' method, where the orbitals are rotated and optimized in an outer loop, with the active space configuration interaction exactly solved at each orbital point in an inner loop \cite{head1988optimization,helgaker2014molecular}.

\subsubsection{Example Numerical Performance of C-DF}

The C-DF approach outlined above was implemented in a simple \texttt{python/numpy} environment. L-BFGS (\texttt{scipy}) with analytical gradients is used to drive the optimization loop in $X_{pk}^{t}$. The matrix exponential needed to form $U_{pk}^{t}$ is evaluated from the complex Hermitian eigendecomposition of $i X_{pk}^{t}$, and the corresponding exponential derivative is evaluated by the Wilcox formula \cite{wilcox1967exponential}. The linear equations used to solve for the optimal $Z_{pk}^{t}$ for a given $U_{pk}^{t}$, e.g., Equation \ref{eq:Zlinear}, are solved explicitly via an eigendecomposition-based Moore-Penrose pseudoinverse with eigenvalue cutoff of $10^{-10}$.

\begin{figure}
\includegraphics[width=0.9\columnwidth]{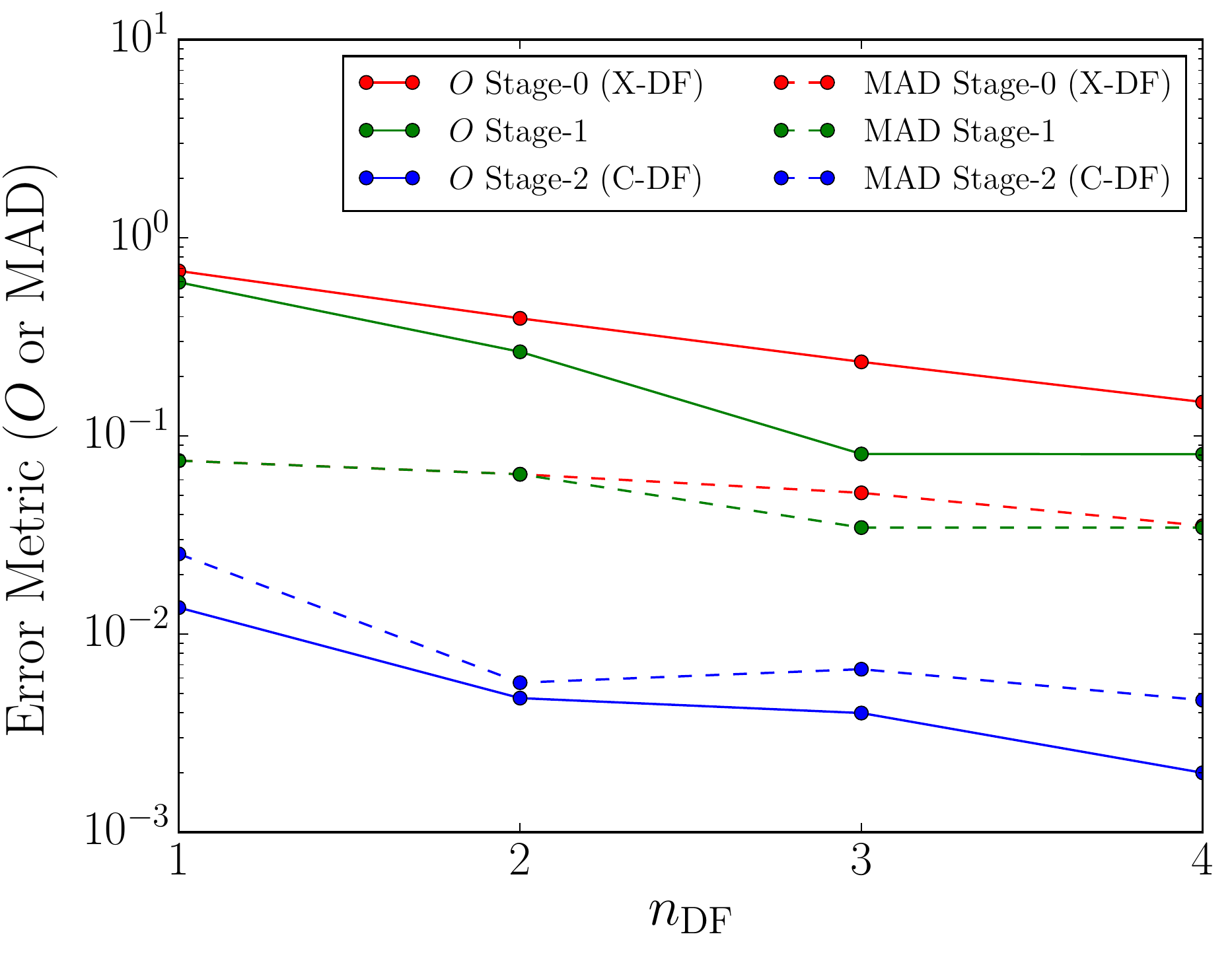}
\caption{
Error metrics for Stage-0 (X-DF), Stage-1, and Stage-2 (C-DF) two-step double factorization approaches. The test case is naphthalene with an $n_{p} = 10$ active space of low-lying $\pi$ and $\pi^*$ orbitals computed at RHF/cc-pVDZ. $O$ is the C-DF least-squares objective function, and MAD is the mean absolute deviation $|\Delta_{pqrs}|_{\infty}$.} 
\label{fig:df-numerical}
\end{figure}

Figure \ref{fig:df-numerical} shows representative performance of the X-DF and C-DF methods. The test case is a 10 orbital active space of the lowest lying $\pi$ and $\pi^*$ orbitals of naphthalene, with the orbitals computed at RHF/cc-pVDZ.  The C-DF objective function $\mathcal{O} (U_{pk}^{t}, Z_{kl}^{t})$ and the maximum absolute deviation (MAD) in the ERI tensor $|\Delta_{pqrs}|_{\infty}$ are plotted as a function of DF rank expansion $n_{\mathrm{DF}}$ for the Stage 0, Stage 1, and Stage 2 outcomes of the C-DF procedure. Stage 0 is the result that would be obtained by the older X-DF approach, while Stage 2 corresponds to a complete C-DF procedure. 

The results are generally straightforward. Using the shortest possible DF factorization $n_{\mathrm{DF}} = 1$, the Stage-0 X-DF obtains a rather coarse representation of the ERI tensor, with objective function value $\delta$ of $\mathcal{O}(10^{0})$ and ERI MAD of $\mathcal{O} (10^{-1})$. Adding more X-DF factors monotonically improves matters, with roughly geometric convergence (as expected from density fitting), however the prefactor is rather slow, and only a 4-fold reduction in objective function value and ERI MAD are achieved by $n_{\mathrm{DF}} = 4$. Moving to Stage-1, one finds that refitting the $Z_{kl}^{t}$ factors with fixed X-DF $U_{kl}^{t}$ factors does not significantly improve matters. However, moving to the full Stage-2 C-DF method, one finds substantial improvement of at least one order of magnitude in $O$ and roughly one order of magnitude in $\delta$ for all $n_{\mathrm{DF}}$. Particularly striking is the fact that it is better to use the coarsest $n_{\mathrm{DF}} = 1$ C-DF factorization than to use the largest $n_{\mathrm{DF}} = 4$ factorization shown here. It should be noted that C-DF is not a panacea in the sense that (1) often a very large number of L-BFGS optimization epochs are required to obtain substantive convergence and (2) in some cases, such as $n_{\mathrm{DF}} = 5$ (not shown on Figure \ref{fig:df-numerical} for clarity), substantive convergence is not obtained at all within $10^{5}$ L-BFGS epochs. This indicates that additional work should be done to improve the convergence behavior of the C-DF optimization procedure. This issue aside, C-DF seems to provide remarkable improvement over X-DF in many cases, particularly including the highly important case of small $n_{\mathrm{DF}}$ DF rank expansion. 

We use C-DF for all QFD cases discussed later in this work.

\subsection{Quantum Filter Diagonalization}

\begin{figure}
\includegraphics{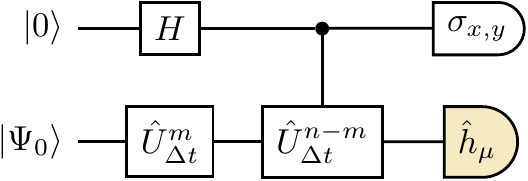}
\caption{Quantum circuits for measuring the QFD matrix element $H_{m,n}$. 
$\hat{h}_\mu$ denotes a term of the Hamiltonian, and $\hat{U}_{\Delta t}$ the circuit implementing 
time evolution under the Hamiltonian for a time step $\Delta t$. The overlap matrix element $S_{m,n}$ 
is measured replacing $\hat{h}_\mu$ with the identity operator.}
\label{fig:qfd_1}
\end{figure}

Quantum filter diagonalization (QFD) is a technique for approximating eigenpairs of a 
Hamiltonian operator $\hat{H}$ \cite{parrish2019quantum,stair2020multireference}. 
It makes use of a set of of time-propagated 
states,
\begin{equation}
\label{eq:qfd_1}
| \Psi_m \rangle = e^{ - i \Delta t m \hat{H}} | \Psi_0 \rangle
= \hat{U}_{\Delta t}^m | \Psi_0 \rangle
\quad,
\end{equation}
as a variational basis for approximate diagonalization of $\hat{H}$.
In \eqref{eq:qfd_1}, $\hat{U}_{\Delta t}$ is the time-evolution operator under the 
Hamiltonian $\hat{H}$ for time $\Delta t$, $m = 0 \dots \nqfd-1$ is an integer number, 
and $| \Psi_0 \rangle$ is a ``guess'' quantum state determined by classical pre-processing that 
can be prepared efficiently by a quantum circuit, such as a Slater determinant. 
A straightforward variant of the method allows for the use of a basis of multiple nonredundant guess states \cite{parrish2019quantum}, but in this work, we always use a single guess state.
Approximations for the eigenstates of $\hat{H}$ are constructed as linear 
combinations of the basis vectors,
\begin{equation}
\label{eq:qfd_2}
| \Phi_I \rangle = \sum_{m=0}^{\nqfd-1} c_{mI} | \Psi_m \rangle
\quad,
\end{equation}
where the coefficients are determined by classically solving the generalized 
eigenvalue equation $H c_{I} = \varepsilon_I  c_I$,
where
\begin{equation}
\label{eq:qfd_3}
S_{m,n} = \langle \Psi_m | \Psi_n \rangle 
\quad,\quad
H_{m,n} = \langle \Psi_m | \hat{H} | \Psi_n \rangle 
\quad.
\end{equation}
The overlap and Hamiltonian matrices $S$ and $H$ are computed using a set of extended Hadamard quantum circuits \cite{aharonov2009polynomial} 
with a single ancilla, illustrated in Fig.~\ref{fig:qfd_1}.

Indeed, it can be easily shown that
\begin{equation}
\langle \Psi_m | \hat{h}_\mu | \Psi_n \rangle 
=
\langle \chi | X \otimes \hat{h}_\mu | \chi \rangle
+ i
\langle \chi | Y \otimes \hat{h}_\mu | \chi \rangle
\quad,
\end{equation}
where
\begin{equation}
| \chi \rangle = \frac{
| 0 \rangle \otimes \hat{U}_{\Delta t}^m | \Psi_0 \rangle 
+ 
| 1 \rangle \otimes \hat{U}_{\Delta t}^{n-m} | \Psi_0 \rangle 
}{\sqrt{2}}
\end{equation}
is the output state of the quantum circuit in Fig.~\ref{fig:qfd_1}.

In general, the exact time-evolution operator $\hat{U}_{\Delta t}$ is not known.
On a digital quantum computer, it can be approximated with accuracy $\delta$ 
at cost scaling up to polynomially with system size, simulation time $\Delta t$ and 
inverse accuracy $\delta^{-1}$ \cite{georgescu2014quantum}. 
A prominent example is the primitive Trotter approximation
\begin{equation}
\label{eq:qfd_4}
\hat{U}_{\Delta t} \simeq \prod_\mu e^{-i \Delta t \hat{h}_\mu}
\quad,\quad
\hat{H} = \sum_\mu \hat{h}_\mu
\quad,
\end{equation}
where, for each operator $\hat{h}_\mu$, the quantum circuits implementing 
$e^{-i \Delta t \hat{h}_\mu}$ is known.

In this work, we introduce a primitive Trotter approximation \eqref{eq:qfd_4} 
into the definition of the basis vectors $| \Psi_m \rangle$. Within such an
approximation, the Toeplitz property $S_{m,n} = S_{0,n-m}$, $H_{m,n} = H_{0,n-m}$  
enjoyed by the exact overlap and Hamiltonian matrix elements \eqref{eq:qfd_2}
is lost, so that the number of quantum circuits to be evaluated scales as 
$\mathcal{O}(\nqfd\,{\,}^{2})$. Nevertheless, the QFD algorithm retains a number of 
desirable features, especially numerical stability and the structure of a 
variational wavefunction Ansatz \cite{parrish2019quantum}.

In the next section, we explore how the C-DF representation of the electron repulsion integral
can be used to economize QFD circuits, and the measurement of QFD matrix elements.

\subsection{Circuit Reduction Strategies}

To describe the circuit reductions allowed by the combination of QFD and DF, we express the Hamiltonian as in Eq.~\eqref{eq:df_hamiltonian},
\begin{equation}
\begin{split}
\hat{H} 
&= 
E_{\mathrm{Ext}}
+
\sum_{k}
f_{kk}^{0}
\hat E_{kk}^{+}(0)
+
\sum_{t}
\sum_{kl}
\frac{ Z_{kl}^{t} }{2}
\hat E_{kk}^{+}(t)
\hat E_{ll}^{+}(t) \\
&= 
E_{\mathrm{Ext}}
+
\changeofbasis{0}{\mathrm{hf}}
\left[ \sum_{k \sigma}
f_{kk}^{0}
\numberop{k,\sigma}
\right]
\changeofbasis{\mathrm{hf}}{0} \\
&+ 
\sum_t
\changeofbasis{t}{\mathrm{hf}}
\left[ \sum_{kl,\sigma\tau}
\frac{ Z_{kl}^{t} }{2}
\numberop{k,\sigma}
\numberop{l,\tau}
\right]
\changeofbasis{\mathrm{hf}}{t}
\end{split}
\end{equation}
Here, $\changeofbasis{\mathrm{hf}}{l}$ denotes a unitary transforming from the Hartree-Fock basis to the 
eigenbasis of the $l$-th term of the Hamiltonian.
The index $l=0$ represents the the 1-body term,
and the indices $l=1 \dots \nlayers$ represent the
terms of the double factorized Hamiltonian.

Under the Jordan-Wigner mapping, number operators 
$\numberop{k,\sigma}$ and products of number 
operators $\numberop{k,\sigma} \numberop{l,\tau}$
take the form
\begin{equation}
\begin{split}
\numberop{k,\sigma}
&=
\frac{\left(1-Z_{k,\sigma}\right)}{2} 
\\
\numberop{k,\sigma}
\numberop{l,\tau}
&=
\frac{\left(1-Z_{k,\sigma}\right)}{2} 
\frac{\left(1-Z_{l,\tau}\right)}{2} 
\end{split}
\end{equation}
As seen, products of number operators contain terms 
that are linear in $Z_{k,\sigma}$. Our goal is to re-organize the Hamiltonian into a new one-body part
and a set of two-body factors involving only products
of the form $Z_{k,\sigma} Z_{l,\tau}$.
To this purpose, we introduce the operators
$\zetaop{k,\sigma} = 1-2 \numberop{k,\sigma}$ and
recall that $\zetaop{k,\sigma}^2 = 1$, we can 
readily recast Eq.~\eqref{eq:df_hamiltonian} 
in the form
\begin{equation}
\begin{split}
\hat{H} 
&= 
E_{\mathrm{Ext}}^\prime
+
\changeofbasis{0}{\mathrm{hf}}^\prime
\left[ \sum_{k \sigma}
f_{kk}^\prime
\,
\zetaop{k,\sigma}
\right]
\changeofbasis{\mathrm{hf}}{0}^\prime \\
&+
\sum_t
\changeofbasis{t}{\mathrm{hf}}
\left[ \sum_{kl,\sigma\tau}^*
\frac{ Z_{kl}^{t} }{8}
\zetaop{k,\sigma}
\zetaop{l,\tau}
\right]
\changeofbasis{\mathrm{hf}}{t}
\end{split}
\end{equation}
where the asterisk denotes summation over 
strings $kl,\sigma\tau$ with $k \neq l$, 
or $k=l$ and $\sigma \neq \tau$, and primes
denote a simple redefinition of the Hamiltonian coefficients and of the unitary transforming
from the Hartree-Fock basis to the eigenbasis
of the one-body part of the Hamiltonian.

In this form, time evolution under the Hamiltonian
for a time step $\Delta t$ can be approximated by
\begin{equation}
\label{eq:df_hamiltonian_circuit}
\hat{U}_{\Delta t}
\simeq 
e^{- i \Delta t E_{\mathrm{Ext}}^\prime}
\,
\changeofbasis{\nlayers}{\mathrm{hf}}
\left[
\prod_t 
\hat{V}_{\mathrm{2b},t} 
\,
\changeofbasis{t-1}{t}
\right]
\hat{V}_{\mathrm{1b}} 
\,
\changeofbasis{\mathrm{hf}}{0}^\prime
\end{equation}
where 
\begin{equation}
\begin{split}
\hat{V}_{\mathrm{1b}} 
&=
\prod_{k,\sigma}
e^{ - i \Delta t f_{kk}^\prime \, \zetaop{k,\sigma}}
\quad, \\
\hat{V}_{\mathrm{2b},t}
&=
\prod_{kl,\sigma\tau}^*
e^{ - i \Delta t \frac{ Z_{kl}^{t} }{8}
\zetaop{k,\sigma}
\zetaop{l,\tau} }
\quad,
\end{split}
\end{equation}
and 
$\changeofbasis{t-1}{t} = \changeofbasis{t-1}{\mathrm{hf}}
\changeofbasis{\mathrm{hf}}{t}$.

In this form, each of the change-of-basis unitaries $\hat{G}$ factors in two identical parts, acting on spin-up and spin-down spin-orbitals respectively. 
Such parts can be compiled into networks of Givens rotations, which can in turn be represented with 
one- and two-qubit gates, as discussed in the 
Appendix.
Furthermore, in the Jordan-Wigner representation,
evolution under the one-body part of the Hamiltonian
can be implemented by a network of single-qubit $\mathsf{Z}$ rotations, and each of the terms
$\exp\left(-i \,\Delta t \, Z^t_{kl} \, \zetaop{k,\sigma} \,
\zetaop{l,\tau} \right)$ can be implemented with
2 \cnot and and single-qubit $\mathsf{Z}$ rotation, as discussed 
in the Appendix.

\begin{figure}
\includegraphics[width=\columnwidth]{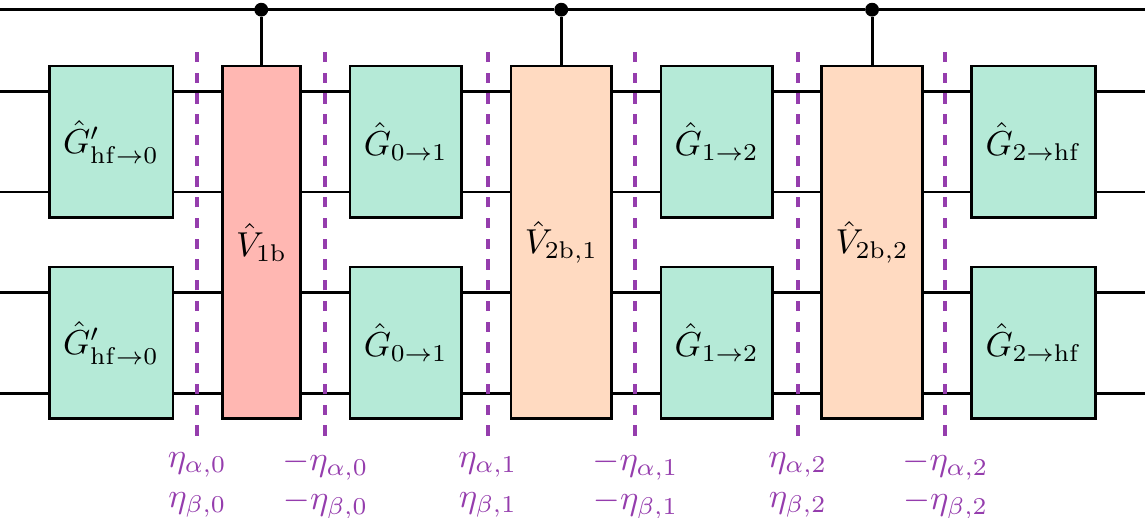}
\caption{Quantum circuit implementing time evolution under the DF Hamiltonian for a time step $\Delta t$. Green, red and orange blocks denote change-of-basis unitaries, evolution under diagonal one-body, and evolution under diagonal two-body operators respectively, and 
violet lines denote echo unitaries parametrized by angles $\eta_k$.}
\label{fig:qfd_2}
\end{figure}

The quantum circuit implementing the controlled 
version of the unitary transformation Eq.~\eqref{eq:df_hamiltonian_circuit} is shown
in Fig.~\ref{fig:qfd_2}.
It is useful to observe that unitaries $\hat{G}$ 
multiply to the identity and thus, as seen in 
Fig.~\ref{fig:qfd_2}, they need not be controlled.
In the Jordan-Wigner representation, each 
controlled $\exp\left( - i \Delta t \, f_{kk}^\prime 
\, \zetaop{k,\sigma} \right)$ can be constructed
with 2 \cnot and 2 single-qubit $\mathsf{Z}$ rotations, whereas each
controlled $\exp\left(-i \,\Delta t \, Z^t_{kl} \, \zetaop{k,\sigma} \,
\zetaop{l,\tau} \right)$ by 
4 \cnot and 2 two-qubit $\mathsf{ZZ}$ rotations.

\begin{figure}
\includegraphics[width=0.4\columnwidth]{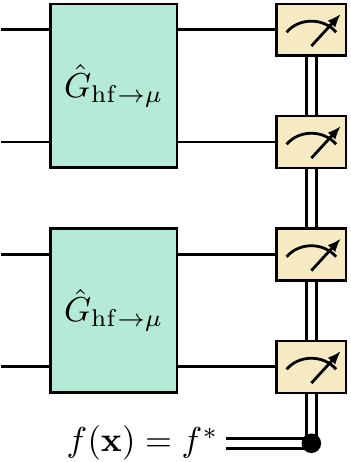}
\caption{Quantum circuit implementing the measurement of a 
term of the C-DF hamiltonian. Green blocks denote change-of-basis unitaries, and yellow meters measurement of single qubits.
Only measurement outcomes ${\bf{x}}$ with correct number of spin-up and spin-down particles are retained in the post-selection algorithm.}
\label{fig:qfd_3}
\end{figure}

\subsection{Error Mitigation Strategies}
\label{sec:error_mitigation}

\subsubsection{Post-Selection}

The structure of the Hamiltonian highlighted in Eq.~\eqref{eq:df_hamiltonian_circuit} allows for 
a simple scheme for measurement post-selection, 
based on enforcing the correct numbers $N_{\alpha}$, 
$N_{\beta}$ of particles for each spin species\cite{huggins2021efficient}. The $\hat{h}_{\mu}$ 
referenced in Fig.~\ref{fig:qfd_1} are represented 
by our operators 
\begin{equation}
\begin{split}
\hat{h}_0 &= \sum_{k \sigma}
f_{kk}^\prime
\zetaop{k,\sigma} 
\quad,
\\
\hat{h}_t &= 
\sum_{kl,\sigma\tau}^*
\frac{ Z_{kl}^{t} }{8}
\zetaop{k,\sigma}
\zetaop{l,\tau}
\quad,\quad t =1 \dots \nlayers \quad.
\end{split}
\end{equation}
As shown in Figure \ref{fig:qfd_3}, we can 
measure each of these operators in their 
respective diagonal basis by first applying 
an appropriate unitary transformation, and
then making a projective measurement in the computational basis. 

The benefit of this measurement scheme is that $N_{\alpha}$ and $N_{\beta}$ are simultaneously diagonalized in each basis. This means that $N_{\alpha}$ and $N_{\beta}$ can be extracted from each measurement shot of each operator in our Hamiltonian. From this information we can discard any shots that return an incorrect number of particles of each spin polarization.

\subsubsection{Echo-sequencing}

In Ref~\cite{tran2021faster} the authors introduce the idea of echo-sequencing each full Trotter step by the inherent symmetries of the Hamiltonian. A standard simulation
\begin{equation}
e^{-it\hat{H}} \approx \prod_j \hat{U}_{\Delta t} \,
\end{equation}
is replaced with a symmetry-protected one,
\begin{equation}
e^{-it\hat{H}} \approx 
\prod_j 
\hat{C}^{\dagger}_j \hat{U}_{\Delta t} \hat{C}_j
\, , \, 
[\hat{C}_j,\hat{H}]=0
\quad,
\end{equation}
resulting in a reduction of the second-order Trotter error.

In this work, the Trotter step takes the form
\begin{equation*}
U_{\mathrm{target}}(\Delta t)
=
\prod_{\mu}e^{-i \Delta t \hat{h}_{\mu}}
\quad,
\end{equation*}
and, to first order in the Schr\"{o}dinger representation, time evolution acts as
\begin{equation*}
\rho(\Delta t)
=
\rho(0)
-
i\Delta t
\sum_{\mu}
\big[ \rho(0) , \hat{h}_{\mu} \big]
\quad.
\end{equation*}
Here, $\big[\hat{N}_{\alpha/\beta},\hat{h}_{\mu}\big]=0$, which provides the opportunity to simultaneously echo both the $\alpha$ and $\beta$ spin sectors for each term $\exp[-i \Delta t \hat{h}_{\mu}]$. Making the assumption 
that the noise exhibited on the hardware, when implementing each time evolved block $\exp[-i \Delta t \hat{h}_{\mu}]$, takes the form
\begin{equation}
U_{\mathrm{exp}}(\Delta t)=\prod_{\mu}e^{-i \Delta t (\hat{h}_{\mu}+\hat{V}_{\mu})}
\quad,
\end{equation}
then, when expanding to first order in the Heisenberg representation, we now have
\begin{equation}
\rho(\Delta t)=\rho(0)-i\Delta t\sum_{\mu}\Big[\rho(0),\hat{h}_{\mu}+\hat{V}_{\mu}\Big]
\quad.
\end{equation}
On the other hand, if we echo with both $\hat{N}_{\alpha}$ and $\hat{N}_{\beta}$ for each term $\exp[-i \Delta t \hat{h}_{\mu}]$, this leaves
\begin{equation}
\begin{split}
U_{\mathrm{echo}}(\Delta t)
&=
\prod_{\mu} 
\hat{C}_\mu
e^{-i \Delta t (\hat{h}_{\mu}+\hat{V}_{\mu})}
\hat{C}_\mu^\dagger \quad, \\
\hat{C}_\mu 
&= 
e^{i \eta_{\alpha,\mu} \hat{N}_{\alpha} } 
e^{i \eta_{\beta,\mu} \hat{N}_{\beta} } \quad, 
\end{split}
\end{equation}
where $\eta_{\alpha/\beta,\mu}$ is a random phase with uniform distribution in the interval $[0,2\pi]$. 
The average of the first-order expansion over the random
phases is
\begin{equation}
\bar{\rho}(\Delta t)
=
\rho(0)
-
i \Delta t
\sum_{\mu}
\Big[\rho(0),
\hat{h}_{\mu}+P^{\dagger}_{N_{\alpha}
,
N_{\beta}}
\hat{V}_{\mu}P_{N_{\alpha},N_{\beta}}
\Big] \quad ,
\end{equation}
where $P_{N_{\alpha},N_{\beta}}$ is the projection operator on the proper symmetry sector. 

Therefore, this echo-sequencing scheme results in the suppression of error terms that couple different symmetry sectors together \cite{bonet2018low,mcardle2019error}. The ability to conduct $L+1$ echoes per Trotter step instead of a single echo should result in a higher capacity for error mitigation, especially when $L$ is large.

Under the Jordan-Wigner mapping the echo terms for each spin species is simply a product of single-qubit $\mathsf{Z}$ rotations, 
which can be implemented with minimal overhead.
The echo scheme can be implemented under the parity mapping as well, 
but this requires applying the operators $\exp(-i \eta (Z_0+Z_0Z_1+Z_1Z_2+...+Z_{N-1}Z_N))$, 
and thus involves a greater overhead. 

\begin{figure}[t]
\includegraphics[width=\columnwidth]{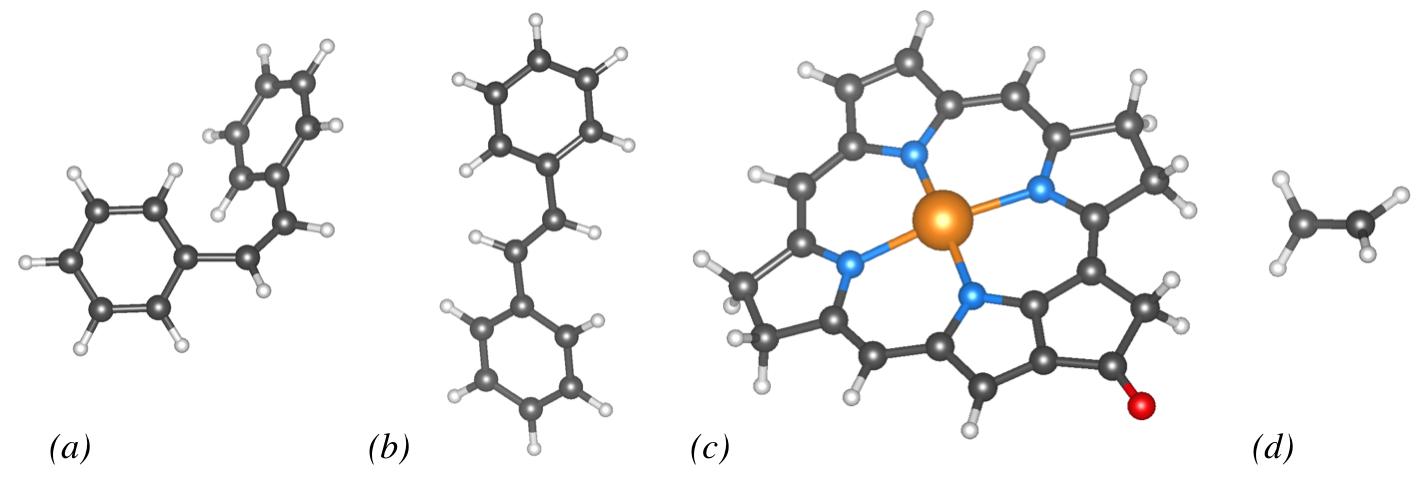}
\caption{Molecular species studied in the present work: 
cis- and trans-stilbene, bacteriochlorophyll $a$ (BChl $a$) 
and ethylene (left to right, a to d).}
\label{fig:res_1}
\end{figure}

\section{Results and Discussion}

\begin{figure}
\includegraphics[width=0.9\columnwidth]{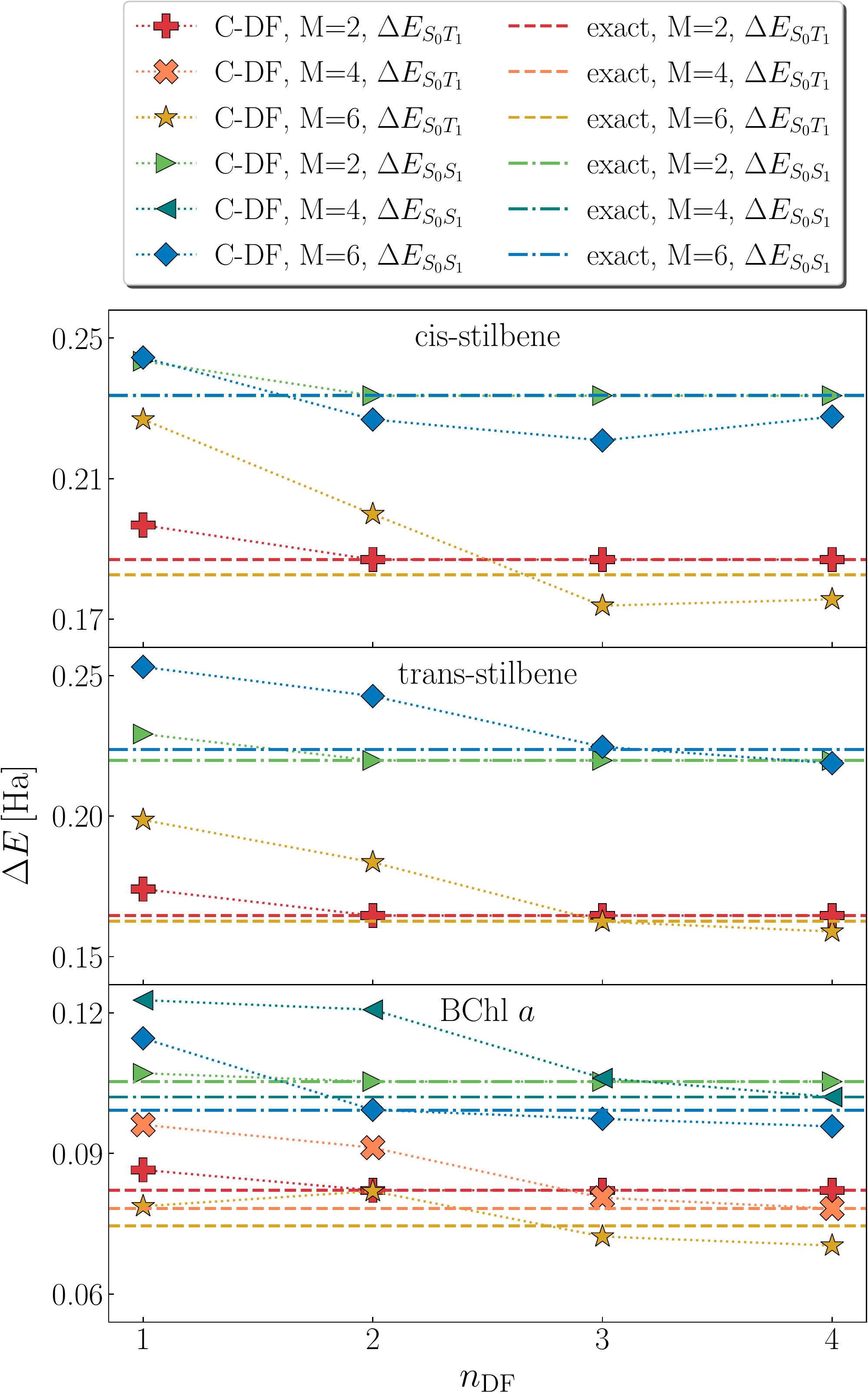}
\caption{
Singlet-triplet $\Delta E_{S_0 T_1}$ and singlet-singlet 
$\Delta E_{S_0 S_1}$ gap from the exact (dashed, dash-dotted lines) 
and the C-DF Hamiltonian (markers) with $\nlayers=1,2,3,4$ 
layers, for cis- and trans-stilbene, and BChl $a$ (top to bottom)
in active spaces of $M=2,4,6$ orbitals (red, orange, yellow and 
green, teal, blue symbols).}
\label{fig:res_2}
\end{figure}

\begin{figure*}[!t]
\includegraphics[width=0.95\textwidth]{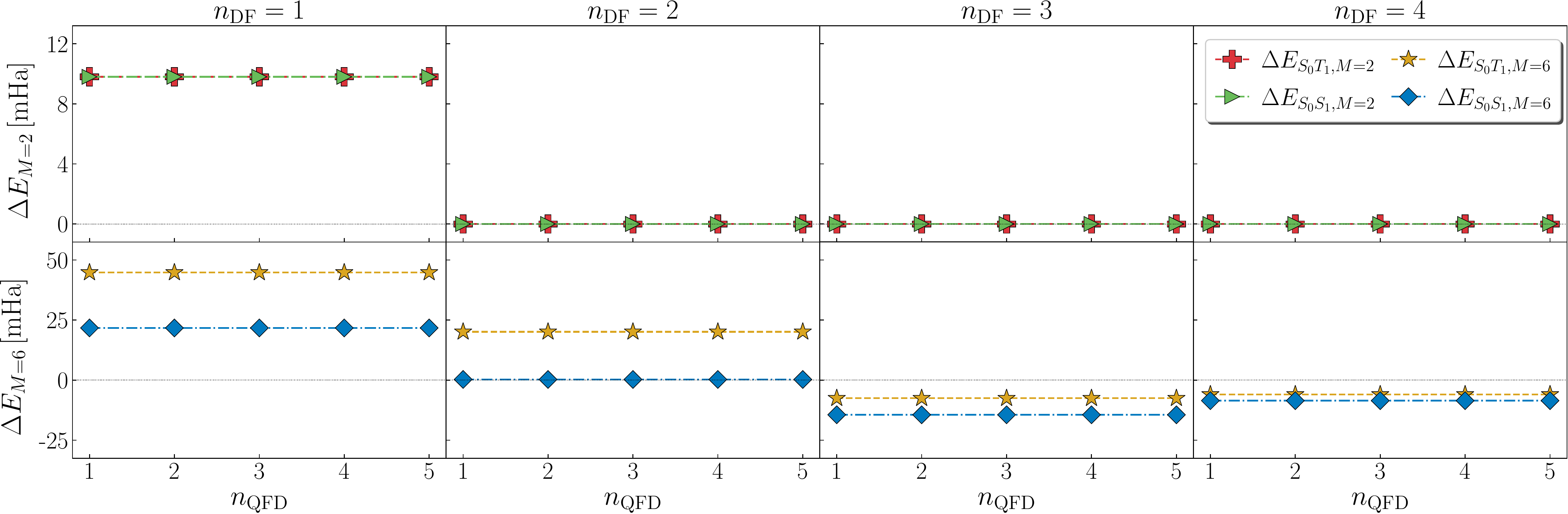}
\caption{Deviation between QFD and FCI singlet-triplet $\Delta E_{S_0 T_1}$ and singlet-singlet $\Delta E_{S_0 S_1}$ gap as a function of the number of QFD steps, for cis-stilbene, in active spaces of $M=2,6$ orbitals (top, bottom) employing $n_{\mathrm{DF}} = 1,2,3,4$ layers (left to right). Calculations use a time step of $\Delta t = 0.1 \,\mathrm{Ha}^{-1}$.}
\label{fig:cis-stilbene_fig2}
\end{figure*}

\begin{figure*}[!t]
\includegraphics[width=0.95\textwidth]{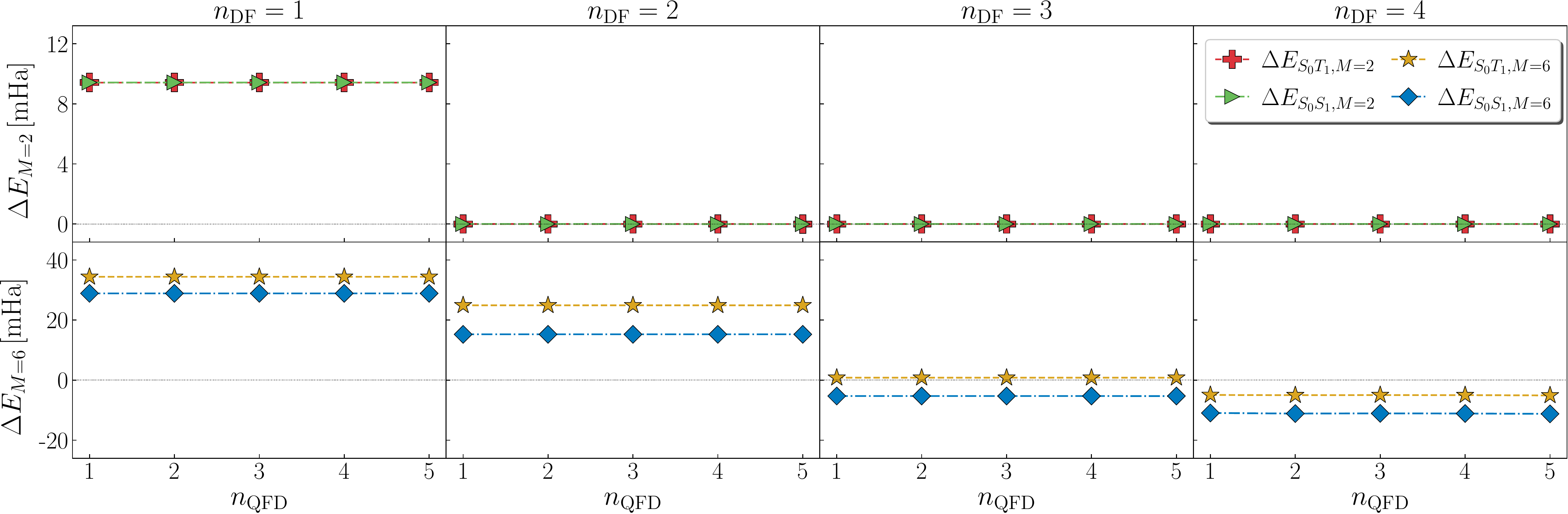}
\caption{Same as Fig.~\ref{fig:cis-stilbene_fig2}, for 
trans-stilbene.}
\label{fig:trans-stilbene_fig2}
\end{figure*}

The strategy for the calculations performed in this work involved 
initial pre-processing by classical quantum chemistry codes on 
conventional computers, to generate optimized Hartree-Fock orbitals 
and matrix elements of the Hamiltonian in active spaces of 2, 4 or 6
orbitals, 
prior to performing computations with quantum simulators or devices.

The chemical species studied in this work are shown in Figure 
\ref{fig:res_1} at geometries listed in the Supplementary Material.
The Hamiltonian is constructed from an active space of restricted Hartree-Fock (RHF) singlet spatial orbitals, computed via the Lightspeed/TeraChem package, for all computations performed herein.

Quantum calculations are performed using IBM's open-source Python 
library for quantum computing, Qiskit \cite{aleksandrowicz2019qiskit}. 
Qiskit provides tools for various tasks such as creating quantum 
circuits, performing simulations, and computations on quantum devices.
We ran our experiments on both the statevector and qasm simulators 
in Qiskit, and performed hardware experiments on 16- and 28-qubit 
devices available through IBM Quantum Experience with quantum volume 
\cite{cross2019QV} of 32, namely, \device{guadalupe}, \device{montreal}
and \device{mumbai} \cite{IBMQDevices}.

\begin{figure*}[!t]
\includegraphics[width=0.95\textwidth]{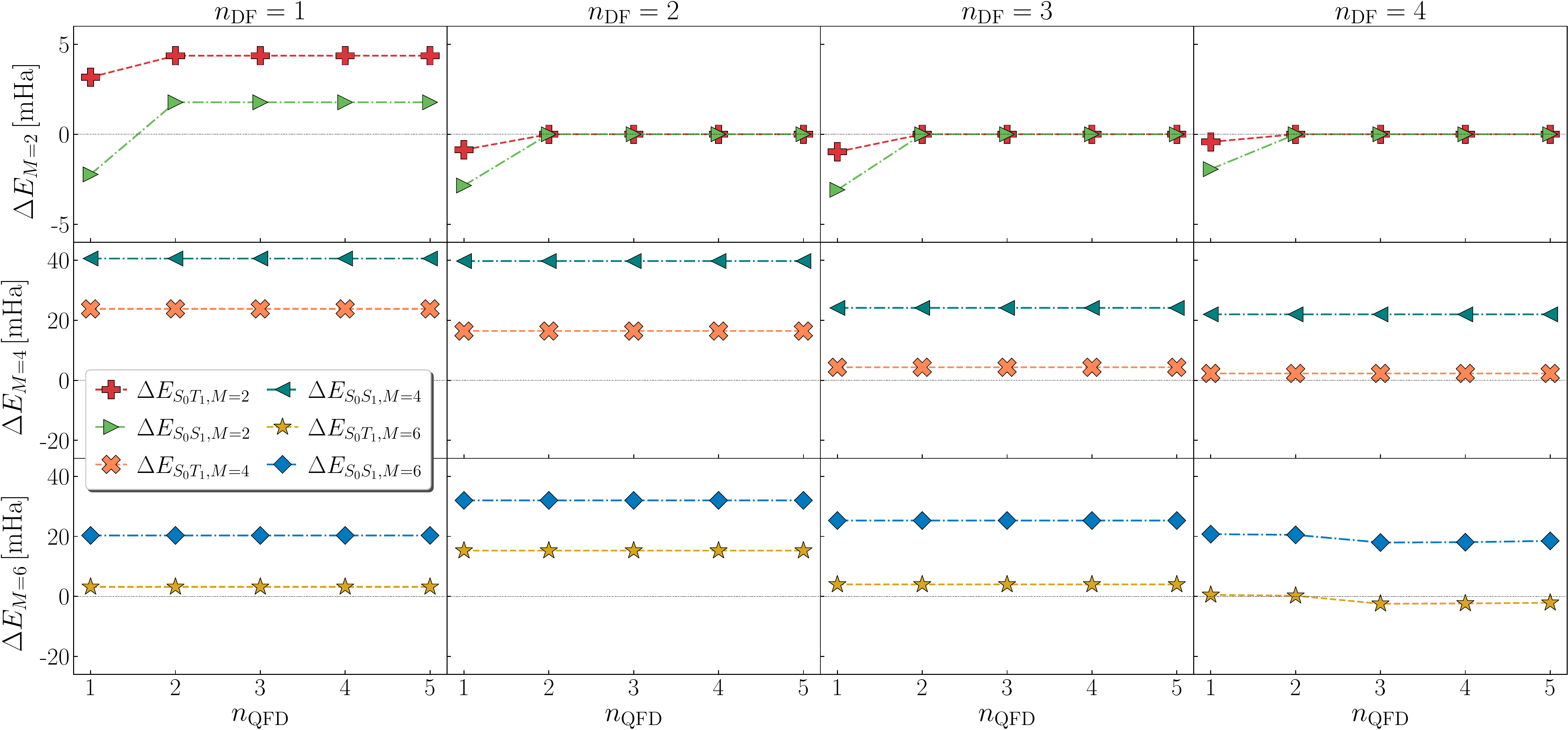}
\caption{Same as Fig.~\ref{fig:cis-stilbene_fig2}, for
BChl $a$. Calculations use a time step of $\Delta t = 0.1 \,\mathrm{Ha}^{-1}$, and active spaces of $M=2,4,6$ (top to bottom) orbitals.}
\label{fig:bchl-a_fig2}
\end{figure*}

\begin{figure*}[!t]
\includegraphics[width=0.95\textwidth]{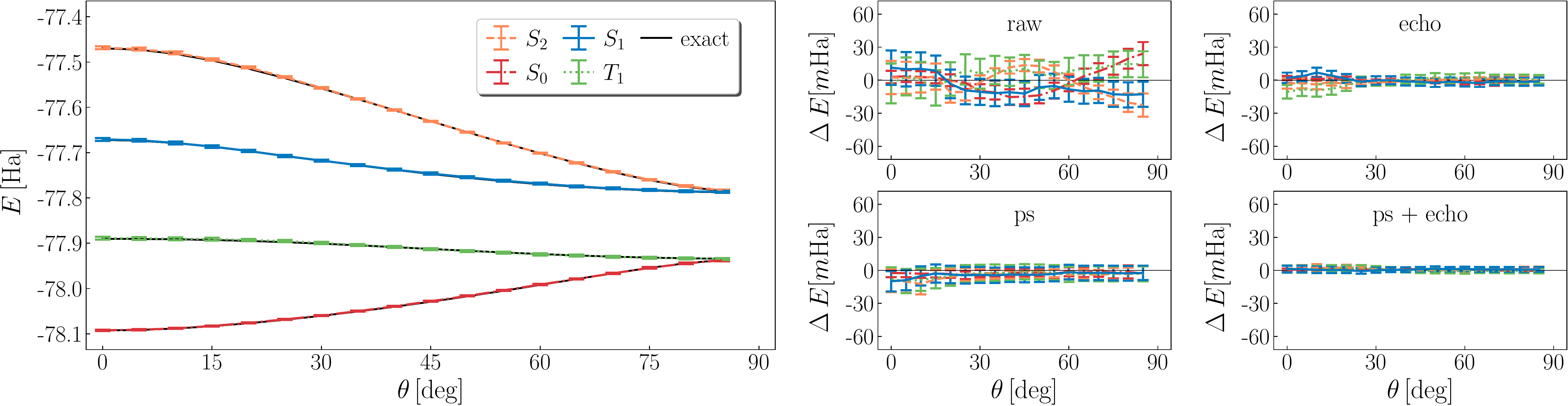}
\caption{Left: exact (black) and QFD (colored) energies as a function of torsion angle for ethylene, using a $M=2$ orbital active space, $n_{\mathrm{DF}}=n_{\mathrm{QFD}}=1$ and $\Delta t = 0.1 \, \mathrm{Ha}^{-1}$. Post-selection and echo-sequencing with $\necho=10$ samples are used, and calculations are carried out on a classical simulator with noise model from \device{montreal}.
Right: differences between QFD and exact energies for calculations with no error mitigation (raw), post-selection (ps), echo-sequencing (echo), and post-selection and echo-sequencing (ps+echo).
}
\label{fig:qasm}
\end{figure*}

\subsection{Classical simulations}

We begin our analysis by computing, in Figure \ref{fig:res_2}, the 
exact singlet-triplet and singlet-singlet gaps, 
$\Delta E_{S_0 T_1}$ and $\Delta E_{S_0 S_1}$ respectively,
for stilbene and BChl $a$. We use the exact and doubly-factorized
Hamiltonian, with $\nlayers = 1,2,3,4$ layers of tensors, with
the purpose of assessing the accuracy of the double factorization.
As seen, for active spaces of $M=2$ orbitals, $\nlayers=2$ layers
are sufficient to obtain an exact representation of the electron
repulsion integral, and thus exact gaps.
For $M=4,6$ orbitals, on the other hand, $\nlayers=4$ layers are
needed to achieve milliHartree accuracy.

In Figures \ref{fig:cis-stilbene_fig2}, \ref{fig:trans-stilbene_fig2}
and \ref{fig:bchl-a_fig2} we perform exact simulations
of the QFD algorithm for cis-, trans-stilbene and BChl $a$ respectively, using active spaces of $M=2$ to $6$ orbitals.
As naturally expected, and in accordance with the observations of Figure \ref{fig:res_2}, increasing the number $\nlayers$ of 
layers in the doubly-factorized representation of the Hamiltonian brings energy differences systematically closer to exact values for the full Hamiltonian.
Indeed, with $\nlayers=1$, deviations between singlet-triplet and singlet-singlet gaps of the exact and doubly-factorized Hamiltonian are of the order of 2 to 40 mHa, and decrease as $\nlayers$ increases.
For all species considered here, the singlet-singlet gap from the doubly-factorized Hamiltonian is closer to the exact value than the singlet-singlet gap, indicating that the latter quantity is more sensitive to approximations in the electron repulsion integral.

For cis- and trans-stilbene, increasing the number $\nsteps$ of time evolution steps in the QFD algorithm has little effect on energy differences. A different behavior is seen in Figure \ref{fig:bchl-a_fig2} for BChl $a$, where energy differences show variations of of the order of a few milliHartree as $\nsteps$ varies. The difference between the behavior of stilbene and BChl $a$ stems from the different point group symmetries
of the geometries studied here ($C_2$ and $C_1$ 
for stilbene and BChl $a$ respectively). Such a difference is particularly visible in the case 
of active spaces with $M=2$ orbitals: for $C_2$-symmetric species, the ground and $S_2$ excited state lie in the $A$ irrep of the $C_2$ symmetry group, and are thus automatically orthogonal to the triplet and $S_1$ excited states, which instead lie in the $B$ irrep. Therefore, a single step of time evolution applied to the Hartree-Fock state ($A$ irrep) or to a configuration with a single HOMO-LUMO excitation ($B$ irrep) is sufficient to completely span the subspaces of $A$ and $B$ symmetric wavefunctions. In the more general case of BChl $a$, 
where such a simplification does not occur, two time evolution steps are needed.

In Figure \ref{fig:qasm}, we perform classical emulations of the QFD algorithm, using a simulator
($\mathsf{qasm}$)
that accounts for statistical uncertainties affecting results of quantum mechanical measurements, and incorporate the effect of various decoherence phenomena through noise models. 
Errors arising from decoherence are mitigated with a combination of the post-selection and echo-sequencing techniques described in the Methods section.

The energies of ground $S_0$ and $T_1$, $S_1$ and $S_2$ excited states are found to be in agreement with exact results across torsion of the C$-$C bond.
The effect of post-selection and echo-sequencing are illustrated in the right part of the figure: raw (i.e. unmitigated) data have deviations from exact results and statistical uncertainties of the order of a few tens of milliHartree. Upon post-selection, both deviations and statistical uncertainties decrease to order 10 milliHartree.

A similar effect is seen when the echo-sequencing technique is applied, which arises because of echo-sampling, and because averaging results over $\necho$ calculations reduces statistical uncertainties by a factor $\necho^{-1/2}$.
Finally, the combination of post-selection and echo-sequencing is seen to reduce statistical uncertainties to 1-2 milliHartree, and deviations between computed and exact results are statistically compatible with zero within such statistical uncertainties.

\subsection{Hardware experiments}

\begin{figure}
\includegraphics[width=0.85\columnwidth]{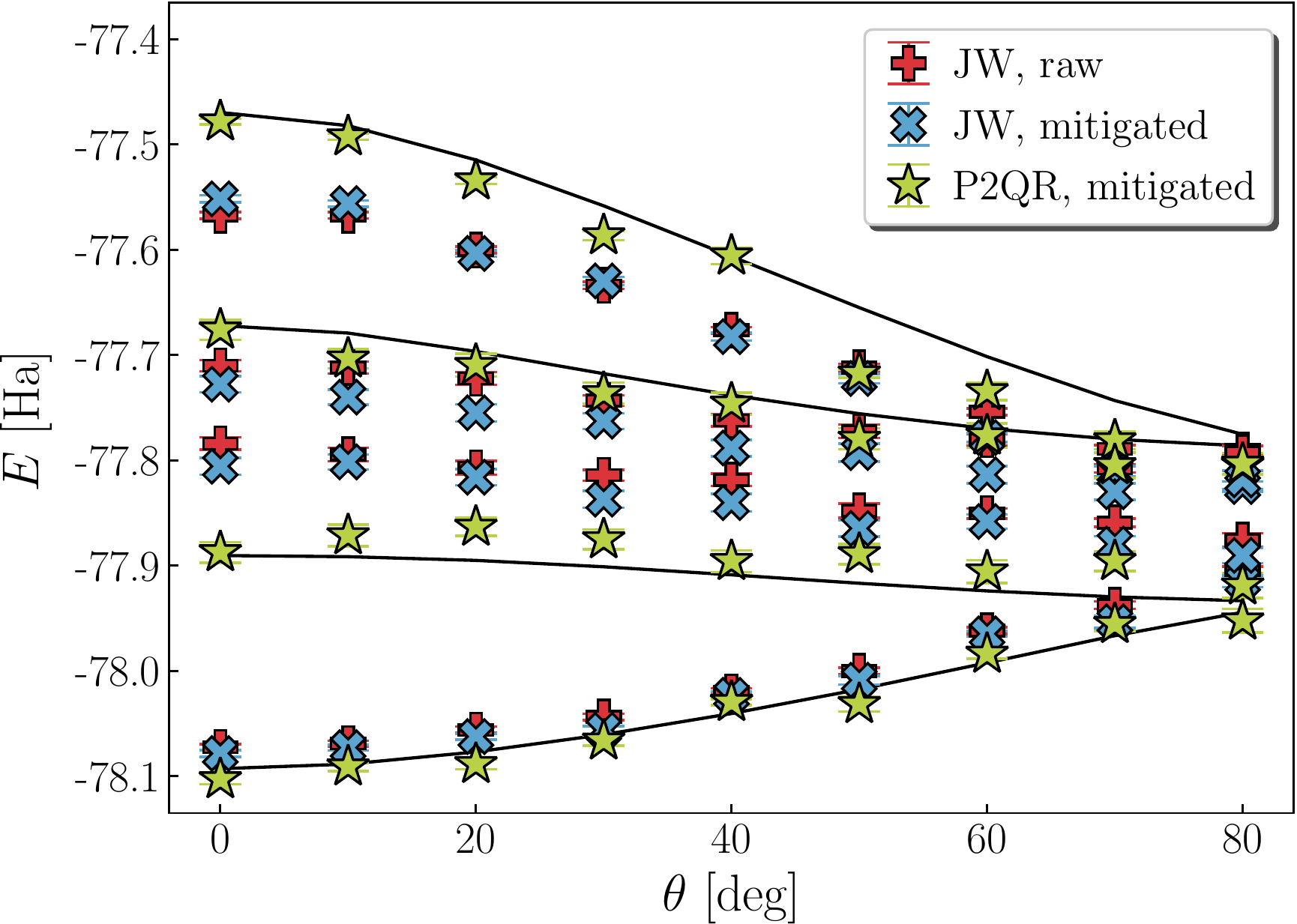}
\caption{Exact (black lines) and computed
(colored points)
nergy spectra for twisted configurations of ethylene. Computations are performed on 
\device{mumbai}, using JW 
(red plus symbols, blue crosses for raw, error-mitigated results) and P2QR
(green stars) representations.}
\label{fig:ethy_curve}
\end{figure}

\begin{figure}
\includegraphics[width=0.85\columnwidth]{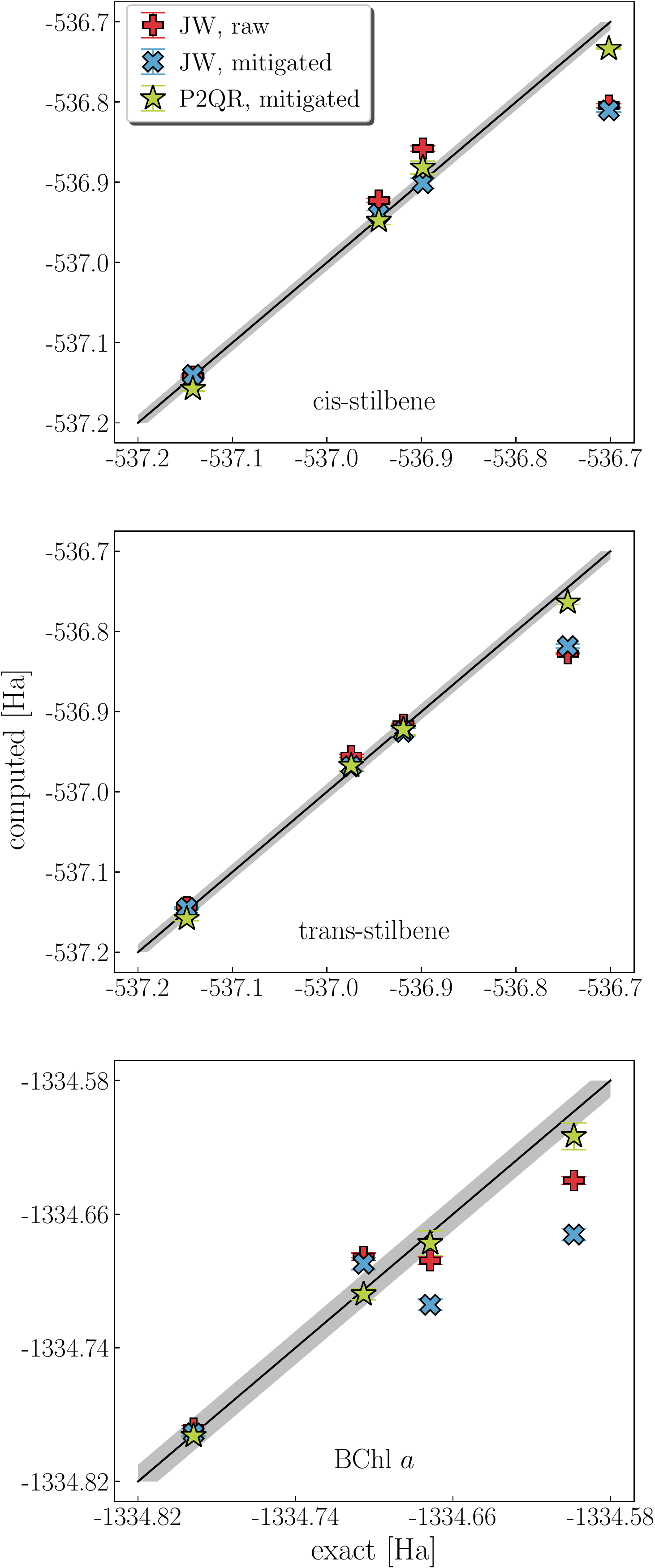}
\caption{Exact and computed energies for cis-stilbene, trans-stilbene and BChl $a$ using Jordan-Wigner representation (red, blue crosses for raw and corrected results) and parity representation with two-qubit reduction (green stars). The gray band has a width of 10 milliHartree. Computations were carried out on
\device{mumbai} and \device{guadalupe}.}
\label{fig:still_bacl}
\end{figure}

All of the hardware experiments are run with a HOMO-LUMO active space, with $\nlayers=1$, and a single time step. We also compare the calculations performed using parity mapping and 2-qubit reduction (P2QR) and Jordan Wigner (JW) which require a total of 3 and 5 qubits respectively. The restriction to a single time step and single C-DF factor is due to constraints on the circuit depth by the inherent noise present in the hardware. 

While JW calculations face a considerable handicap, compared against P2QR, in terms of qubit and total gate count, JW mapping generally has favorable gate count for time evolution of electronic structure Hamiltonian as the number of orbitals increases. As such, our 5-qubit JW results serve as an important benchmark on the path to simulating larger systems.

We also employ the use of two reference states to extract the full spectra containing 3 singlets and a triplet state in this active space. This requires a total of 24 circuits (2 reference states each require measurements of 3 matrix elements, each of which requires the real and imaginary parts of the 1-body and 2-body parts of the Hamiltonian) per geometry and each circuit is measured with $8\times10^3$ shots. Measurement error mitigation is also employed, and calibrated using $8\times10^3$ shots.

In Figure~\ref{fig:ethy_curve}, we explore the spectra of ethylene, as a function of torsion angle. Unsurprisingly, the P2QR results outperform the JW results, but they fail to accurately reflect the proper excited state energies when twisted more than $40^{\circ}$. Within the P2QR HOMO-LUMO active space, the number of particles in each spin species is automatically conserved so our error mitigation strategies add no benefit. %Due to time restraints the echo sequences were not implemented in these experiments.   

The accuracy of JW results follow a similar trend with the torsion angle, but induce larger systematic bias, as it requires deeper circuits. Implementing the post-selection scheme within JW reduces the bias on the ground- and lowest excited-state energies, but increases the bias on the other excited energies.  

In Figure \ref{fig:still_bacl}, we report results for cis-stilbene, trans-stilbene and BChl $a$. Most of the P2QR results are statistically compatible with exact energies, with slightly larger hardware errors on the highest energy in each of these experiments. As expected, the JW experiments yield larger systematic errors for the high energy states. In the cis-stilbene plots it is clear that post-selection provides a substantial correction to the first and second excited state. A similar effect is observed in the trans-stilbene case. 

Conversely, in the BChl $a$ result, post-selection mildly reduced the systematic error for the first excited state, but simultaneously shifted the second and third excited states further from their exact values. Given that there is a general trend for post-selection to decrease the energies it is likely that typical errors generate transitions to particle sectors with larger energies.

This still leaves open the source of error for the large bias in some of these data points. We suspect that noise on the ancilla qubit may dramatically affect the results, but leave a rigorous investigation of noise sources to future work. 

\section{Summary and Outlook}

In this work, we have considered a number of related techniques that can be stacked to substantially lower the quantum resources required to perform accurate computations of low-lying spectra of electronic Hamiltonians. The most straightforward step is the merger of the QFD approach with the low rank DF representation of the electronic Hamiltonian, which provides considerable reductions of both the circuit size needed for QFD time propagation and the number of measurements needed for Hamiltonian expectation value calculation. This merger of QFD and DF is further accelerated by intrinsic reductions in the required DF rank expansion afforded by moving from an explicit DF (X-DF) representation to a compressed DF (C-DF) representation. 

One interesting point that was approached but not fully solved within this work involves the positioning of double factorization between density fitting and tensor hypercontraction. Density fitting \cite{
Whitten:1973:4496,
Dunlap:1977:81,
Dunlap:1979:3396,
Feyereisen:1993:359,
Komornicki:1993:1398,
Vahtras:1993:514,
Rendell:1994:400,
Kendall:1997:158,
Weigend:2002:4285} and the closely-related Cholesky decomposition approach for ERIs \cite{
Beebe:1977:683,
Roeggen:1986:154,
Koch:2003:9481,
Aquilante:2007:194106,
Aquilante:2009:154107} reduces the rank-4 ERI tensor to a product of 2 rank-3 tensors $(pq|rs) \approx \sum_{A} L_{pq}^{A} L_{rs}^{A}$, where the auxiliary index size $n_{A}$ is found to scale linearly in $n_{p}$. Tensor hypercontraction \cite{hohenstein2012tensor,parrish2012tensor,parrish2013exact} reduces the ERI tensor to a product of 5 rank-2 tensors, with a structure $(pq|rs) \approx \sum_{kl} X_{pk} X_{qk} Z_{kl} X_{rl} X_{sl}$ that is very similar to double factorization, but without the requirement that the leaf tensors $X_{pk}$ are orthogonal or square (i.e., $n_k$ may be different from, and usually larger than $n_{p}$, though is found to scale linearly in $n_p$). The non-orthogonality of tensor hypercontraction appears to be quite problematic for quantum algorithms, e.g. as evidenced by the need for quantum signal processing approaches in a recent approach for the adoption of tensor hypercontraction into quantum algorithms by the Google team \cite{lee2020even}.  Explicit double factorization reduces the ERI tensor to an $n_{\mathrm{DF}}$-depth sum over unitary tensor hypercontractions, each indexed by $t$, i.e., $(pq|rs) \approx \sum_{tkl} U_{pk}^{t} U_{qk}^{t} Z_{kl}^{t} U_{rl}^{t} U_{sl}^{t}$ with the size of $n_{\mathrm{DF}}$ scaling linearly in $n_p$. The unitary nature of the leaf tensors $U_{pk}^{t}$ in double factorization makes the approach immediately amenable to implementation within quantum algorithms. However, explicit double factorization retains the rank-3 information content and cost of density fitting, rather than the rank-2 information content and cost of tensor hypercontraction. It seems incongruous to us that simply constraining the tensor hypercontraction factorization to use unitary factors to facilitate deployment within quantum algorithms should cause a rise in the information content from rank-2 to rank-3. This motivated our development of the compressed double factorization approach in this work as a pragmatic attempt to reduce the information content in the double factorization approach. Substantial numerical gains were demonstrated, but it is not clear if the resulting method achieves the constant $n_{\mathrm{DF}}$ depth required to obtain rank-2 information content. More work must be done to pursue an analog to the analytical exact tensor hypercontraction result \cite{parrish2013exact} (which rigorously demonstrated the rank-2 information content of tensor hypercontraction) in closed basis sets, and to extend compressed double factorization to a more reliable and practical method for non-closed basis sets. It is also worth noting that C-DF or extensions thereof can easily be deployed in other quantum algorithms besides QFD, e.g. the variational quantum eigensolver and quantum phase estimation.

We also investigate methods for circuit reduction and error mitigation to improve performance on noisy quantum hardware. The circuit reduction is performed by re-organizing our Hamiltonian so that the 2-body DF terms only contain quadratic Pauli-Z terms ($Z_{k,\sigma}Z_{l,\tau}$) while simultaneously defining a new 1-body term. When performing controlled time evolution this process results in a saving of ($2\times N\times \nlayers$) \cnot{s}, where $N$ is the number of qubits and $\nlayers$ is the number of double-factorized terms in the Hamiltonian. 

In our measurement scheme we rotate to a diagonal basis of each factor of the double factorized C-DF Hamiltonian as shown in Figure~\ref{fig:qfd_3}. This enables us to post-select only the results with the proper number of particles in each spin species. An extra layer of error mitigation can be employed by echoing with $\exp(-i \eta_{k,\sigma} \hat{N}_{\sigma})$ as shown in Figure~\ref{fig:qfd_2}. The combined benefit of both these mitigation strategies is demonstrated with the $\mathsf{qasm}$ noise simulation of the ethylene in Figure~\ref{fig:qasm}. As the depth of these circuits increase, and the hardware noise is constrained to modest level, the echo self-averaging effect can enable substantial error reduction within each individual instance. At shorter depths, it is important to average over random instances to smooth out the results. There are many other error mitigation strategies that can be implementing for these calculation. The benefit to the post-selection and echo-sequencing schemes we present is that they naturally fit into the structure of these circuits with minimal resource overhead.

The methods presented in this work were also implemented on IBM's quantum devises with calculations of energy spectra for twisted ethylene, (cis/trans)-stilbene, and BChl $a$. As expected, calculation using P2QR outperformed results that used JW mapping. The JW results were still able to modestly reproduce the proper energy spectra and should be seen as a benchmark for moving to larger systems. 

The post-selection scheme produced a modest improvement in the JW results and we still need to experiment with the echo sequences. It is clear that the noise simulated backend does not faithfully emulate the hardware noise in our experiments. This is readily seen when comparing the the ethylene curves in Figure~\ref{fig:qasm} and Figure~\ref{fig:ethy_curve}. Given that the results are highly sensitive to noise on the ancilla qubit it will be important to investigate these noise source further and determine how well they can mitigated. 

\textbf{Data Availability:} Molecular structures and electronic Hamiltonian matrix elements are available from the authors upon reasonable request.

\textbf{Acknowledgements:} The QC Ware effort in this work was supported by the U. S. Department of Energy,
Office of Science, Basic Energy Sciences, Chemical Sciences, Geosciences
and Biosciences Division.

\textbf{Conflict of Interest:} RMP owns stock/options in QC Ware Corp.

\section{Appendix: Example Circuits}

\begin{figure}[b!]
\includegraphics[width=\columnwidth]{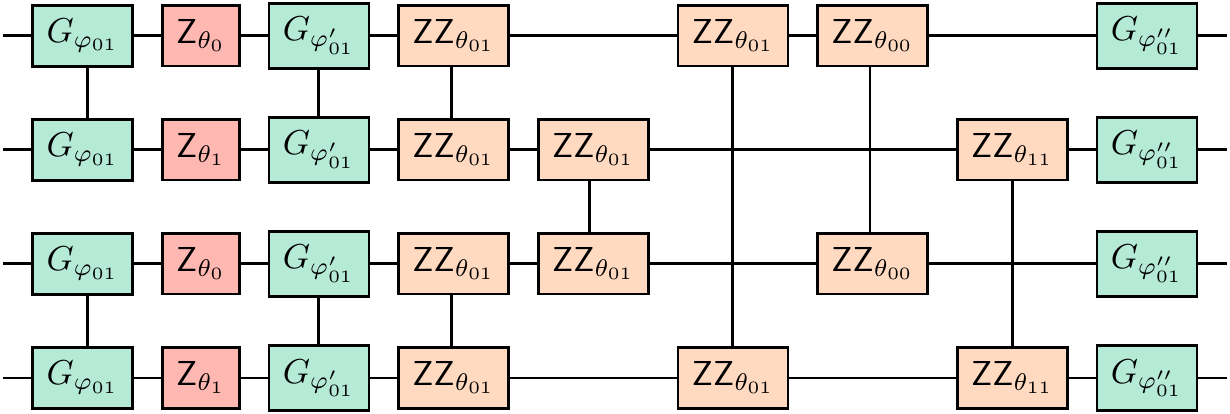}
\caption{Quantum circuit implementing a single step of time evolution under a Hamiltonian with $\nlayers=1$ layers acting on $M=2$ spatial orbitals in a Jordan-Wigner representation. Green, red, and orange blocks denote Givens rotations implementing basis changes, single-qubit $Z$ rotations, and two-qubit $\mathsf{Z}$ rotations respectively.}
\label{fig:circuit_a}
\end{figure}

In this Section, we describe the detailed structure
of the QFD circuits sketched in Figures \ref{fig:qfd_2} and \ref{fig:qfd_3},
focusing on a system of electrons in $M=2$ spatial orbitals, and using the Jordan-Wigner representation.
In Figure \ref{fig:circuit_a}, we show the quantum
circuit corresponding to a step of time evolution
under the Hamiltonian, using a C-DF approximation
of the ERI tensor with $\nlayers = 1$ layers. 

\begin{figure}[b!]
\includegraphics[width=0.8\columnwidth]{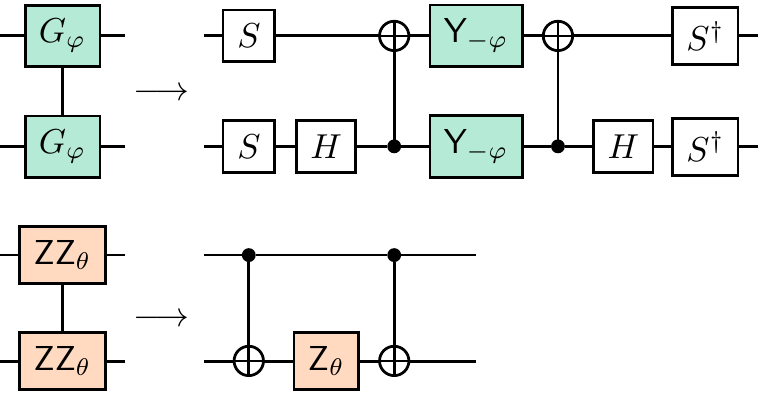}
\caption{Implementation of a Givens rotation (top) and 
of a two-qubit $Z$ rotation (bottom) with \cnot gates 
and single-qubit rotations.}
\label{fig:circuit_b}
\end{figure}

The matrices $\changeofbasis{\mathrm{hf}}{t}$ and $\changeofbasis{t}{t^\prime}$, 
connecting eigenbases of the Fock and C-DF operators, 
are represented as products of Givens transformations 
with standard linear algebra techniques \cite{kivlichan2018quantum,motta2018low}.
Givens transformations correspond to operators of the form
\begin{equation}
\label{eq:givens_operator}
\hat{G}_{r-1,r}(\varphi_{r-1,r}) 
= 
\prod_\sigma 
e^{ - \varphi_{r-1,r} 
\left( 
\hat{a}^\dagger_{r-1 \sigma} \hat{a}_{r \sigma} 
- 
\hat{a}^\dagger_{r \sigma} \hat{a}_{r-1 \sigma} 
 \right) } 
\;,
\end{equation}
which, in a Jordan-Wigner representation, 
are represented by two quantum circuits 
(one for spin-$\alpha$ and one for spin-$\beta$ particles), 
each acting on 2 qubits and comprising 2 \cnot gates, 
as shown in the upper portion of Figure \ref{fig:circuit_b}.


The representation of Givens transformation as
second-quantization operators, Eq.~\eqref{eq:givens_operator},
is also the starting point for deriving the corresponding 
quantum circuit under other representations of fermionic 
degrees of freedom with qubits 
(e.g. parity, Bravyi-Kitaev \cite{bravyi2002fermionic}), 
as well as in presence of qubit reduction techniques.

The one-body part of the Hamiltonian, on the other 
hand, is represented by the diagonal operator
\begin{equation}
\hat{V}_{\mathrm{1b}} = e^{-i \Delta t \sum_{k \sigma} f^\prime_{kk} \zetaop{k\sigma} }
= \prod_{k \sigma} e^{-i \Delta t f^\prime_{kk} \zetaop{k\sigma} }
\quad.
\end{equation}
Since under Jordan-Wigner representation $\zetaop{k\uparrow} \to Z_{k}$ and $\zetaop{k\downarrow} \to Z_{k+M}$,
the operation $\hat{V}_{\mathrm{1b}}$ is implemented by a product
of $2M$ single-qubit $\mathsf{Z}$ rotations with angles 
$\theta_{k} = - 2 \Delta t f^\prime_{kk}$,
\begin{equation}
\hat{V}_{\mathrm{1b}} \to 
\prod_{k \sigma}
\mathsf{Z}_{\theta_{k},k \sigma}
\quad.
\end{equation}
Such single-qubit operations
are shown as red blocks in \ref{fig:circuit_a}.
For the same reason, 
the terms describing the two-body part of the Hamiltonian,
\begin{equation}
\hat{V}_{\mathrm{2b,t}}
=
\prod^*_{kl,\sigma\tau}
e^{ -i \frac{\Delta t}{2} Z^t_{kl} \zetaop{k\sigma} \zetaop{l\tau} }
\end{equation}
are implemented, in a Jordan-Wigner representation, 
by a product of $\mathcal{O}(M^2)$ two-qubit $ZZ$
rotations with angles $\theta_{kl} = \Delta t Z^t_{kl}$,
\begin{equation}
\hat{V}_{\mathrm{2b,t}}
\to
\prod^*_{kl,\sigma\tau}
\mathsf{ZZ}_{\theta_{kl},k\sigma,l\tau}
\quad.
\end{equation}
Such two-qubit transformations, shown as orange 
blocks in \ref{fig:circuit_a}, are exponentials
of the operator $Z \otimes Z$, 
and can thus be compiled into a product of two 
\cnot transformations and a single-qubit $\mathsf{Z}$ 
rotation, as shown in the lower portion of 
Figure \ref{fig:circuit_b}.

It is useful to remark that the controlled 
version of $\hat{V}_{\mathrm{2b,t}}$ only requires
to control the single-qubit $\mathsf{Z}$ rotation, and not
the two \cnot operations. Furthermore, a network of
SWAP gates can be used to ensure that all $\mathsf{ZZ}$ and 
controlled $\mathsf{ZZ}$ rotations act on adjacent qubits 
(assuming linear chip topology) \cite{kivlichan2018quantum}.

\bibliographystyle{apsrev4-2}
\bibliography{qfd,refs,jrncodes}

\begin{thebibliography}{54}%
\makeatletter
\providecommand \@ifxundefined [1]{%
 \@ifx{#1\undefined}
}%
\providecommand \@ifnum [1]{%
 \ifnum #1\expandafter \@firstoftwo
 \else \expandafter \@secondoftwo
 \fi
}%
\providecommand \@ifx [1]{%
 \ifx #1\expandafter \@firstoftwo
 \else \expandafter \@secondoftwo
 \fi
}%
\providecommand \natexlab [1]{#1}%
\providecommand \enquote  [1]{``#1''}%
\providecommand \bibnamefont  [1]{#1}%
\providecommand \bibfnamefont [1]{#1}%
\providecommand \citenamefont [1]{#1}%
\providecommand \href@noop [0]{\@secondoftwo}%
\providecommand \href [0]{\begingroup \@sanitize@url \@href}%
\providecommand \@href[1]{\@@startlink{#1}\@@href}%
\providecommand \@@href[1]{\endgroup#1\@@endlink}%
\providecommand \@sanitize@url [0]{\catcode `\\12\catcode `\$12\catcode
  `\&12\catcode `\#12\catcode `\^12\catcode `\_12\catcode `\%12\relax}%
\providecommand \@@startlink[1]{}%
\providecommand \@@endlink[0]{}%
\providecommand \url  [0]{\begingroup\@sanitize@url \@url }%
\providecommand \@url [1]{\endgroup\@href {#1}{\urlprefix }}%
\providecommand \urlprefix  [0]{URL }%
\providecommand \Eprint [0]{\href }%
\providecommand \doibase [0]{https://doi.org/}%
\providecommand \selectlanguage [0]{\@gobble}%
\providecommand \bibinfo  [0]{\@secondoftwo}%
\providecommand \bibfield  [0]{\@secondoftwo}%
\providecommand \translation [1]{[#1]}%
\providecommand \BibitemOpen [0]{}%
\providecommand \bibitemStop [0]{}%
\providecommand \bibitemNoStop [0]{.\EOS\space}%
\providecommand \EOS [0]{\spacefactor3000\relax}%
\providecommand \BibitemShut  [1]{\csname bibitem#1\endcsname}%
\let\auto@bib@innerbib\@empty
%</preamble>
\bibitem [{\citenamefont {Kempe}\ \emph {et~al.}(2006)\citenamefont {Kempe},
  \citenamefont {Kitaev},\ and\ \citenamefont {Regev}}]{kempe2006complexity}%
  \BibitemOpen
  \bibfield  {author} {\bibinfo {author} {\bibfnamefont {J.}~\bibnamefont
  {Kempe}}, \bibinfo {author} {\bibfnamefont {A.}~\bibnamefont {Kitaev}},\ and\
  \bibinfo {author} {\bibfnamefont {O.}~\bibnamefont {Regev}},\ }\href
  {https://link.springer.com/chapter/10.1007/978-3-540-30538-5_31} {\bibfield
  {journal} {\bibinfo  {journal} {SIAM J. Comput}\ }\textbf {\bibinfo {volume}
  {35}},\ \bibinfo {pages} {1070} (\bibinfo {year} {2006})}\BibitemShut
  {NoStop}%
\bibitem [{\citenamefont {Georgescu}\ \emph {et~al.}(2014)\citenamefont
  {Georgescu}, \citenamefont {Ashhab},\ and\ \citenamefont
  {Nori}}]{georgescu2014quantum}%
  \BibitemOpen
  \bibfield  {author} {\bibinfo {author} {\bibfnamefont {I.~M.}\ \bibnamefont
  {Georgescu}}, \bibinfo {author} {\bibfnamefont {S.}~\bibnamefont {Ashhab}},\
  and\ \bibinfo {author} {\bibfnamefont {F.}~\bibnamefont {Nori}},\ }\href
  {https://dx.doi.org/10.1103/RevModPhys.86.153} {\bibfield  {journal}
  {\bibinfo  {journal} {Rev. Mod. Phys.}\ }\textbf {\bibinfo {volume} {86}},\
  \bibinfo {pages} {153} (\bibinfo {year} {2014})}\BibitemShut {NoStop}%
\bibitem [{\citenamefont {Parrish}\ and\ \citenamefont
  {McMahon}(2019)}]{parrish2019quantum}%
  \BibitemOpen
  \bibfield  {author} {\bibinfo {author} {\bibfnamefont {R.~M.}\ \bibnamefont
  {Parrish}}\ and\ \bibinfo {author} {\bibfnamefont {P.~L.}\ \bibnamefont
  {McMahon}},\ }\href {https://arxiv.org/abs/1909.08925} {\bibfield  {journal}
  {\bibinfo  {journal} {arXiv:1909.08925}\ } (\bibinfo {year}
  {2019})}\BibitemShut {NoStop}%
\bibitem [{\citenamefont {Stair}\ \emph {et~al.}(2020)\citenamefont {Stair},
  \citenamefont {Huang},\ and\ \citenamefont
  {Evangelista}}]{stair2020multireference}%
  \BibitemOpen
  \bibfield  {author} {\bibinfo {author} {\bibfnamefont {N.~H.}\ \bibnamefont
  {Stair}}, \bibinfo {author} {\bibfnamefont {R.}~\bibnamefont {Huang}},\ and\
  \bibinfo {author} {\bibfnamefont {F.~A.}\ \bibnamefont {Evangelista}},\
  }\href {https://pubs.acs.org/doi/10.1021/acs.jctc.9b01125} {\bibfield
  {journal} {\bibinfo  {journal} {J. Chem. Theory Comput}\ }\textbf {\bibinfo
  {volume} {16}},\ \bibinfo {pages} {2236} (\bibinfo {year}
  {2020})}\BibitemShut {NoStop}%
\bibitem [{\citenamefont {Klymko}\ \emph {et~al.}(2021)\citenamefont {Klymko},
  \citenamefont {Mejuto-Zaera}, \citenamefont {Cotton}, \citenamefont
  {Wudarski}, \citenamefont {Urbanek}, \citenamefont {Hait}, \citenamefont
  {Head-Gordon}, \citenamefont {Whaley}, \citenamefont {Moussa}, \citenamefont
  {Wiebe} \emph {et~al.}}]{klymko2021real}%
  \BibitemOpen
  \bibfield  {author} {\bibinfo {author} {\bibfnamefont {K.}~\bibnamefont
  {Klymko}}, \bibinfo {author} {\bibfnamefont {C.}~\bibnamefont
  {Mejuto-Zaera}}, \bibinfo {author} {\bibfnamefont {S.~J.}\ \bibnamefont
  {Cotton}}, \bibinfo {author} {\bibfnamefont {F.}~\bibnamefont {Wudarski}},
  \bibinfo {author} {\bibfnamefont {M.}~\bibnamefont {Urbanek}}, \bibinfo
  {author} {\bibfnamefont {D.}~\bibnamefont {Hait}}, \bibinfo {author}
  {\bibfnamefont {M.}~\bibnamefont {Head-Gordon}}, \bibinfo {author}
  {\bibfnamefont {K.~B.}\ \bibnamefont {Whaley}}, \bibinfo {author}
  {\bibfnamefont {J.}~\bibnamefont {Moussa}}, \bibinfo {author} {\bibfnamefont
  {N.}~\bibnamefont {Wiebe}}, \emph {et~al.},\ }\href
  {https://arxiv.org/pdf/2103.08563.pdf} {\bibfield  {journal} {\bibinfo
  {journal} {arXiv:2103.08563}\ } (\bibinfo {year} {2021})}\BibitemShut
  {NoStop}%
\bibitem [{\citenamefont {Huggins}\ \emph {et~al.}(2020)\citenamefont
  {Huggins}, \citenamefont {Lee}, \citenamefont {Baek}, \citenamefont
  {O’Gorman},\ and\ \citenamefont {Whaley}}]{huggins2020non}%
  \BibitemOpen
  \bibfield  {author} {\bibinfo {author} {\bibfnamefont {W.~J.}\ \bibnamefont
  {Huggins}}, \bibinfo {author} {\bibfnamefont {J.}~\bibnamefont {Lee}},
  \bibinfo {author} {\bibfnamefont {U.}~\bibnamefont {Baek}}, \bibinfo {author}
  {\bibfnamefont {B.}~\bibnamefont {O’Gorman}},\ and\ \bibinfo {author}
  {\bibfnamefont {K.~B.}\ \bibnamefont {Whaley}},\ }\href
  {https://iopscience.iop.org/article/10.1088/1367-2630/ab867b} {\bibfield
  {journal} {\bibinfo  {journal} {New J. Phys}\ }\textbf {\bibinfo {volume}
  {22}},\ \bibinfo {pages} {073009} (\bibinfo {year} {2020})}\BibitemShut
  {NoStop}%
\bibitem [{\citenamefont {Neuhauser}(1990)}]{neuhauser1990bound}%
  \BibitemOpen
  \bibfield  {author} {\bibinfo {author} {\bibfnamefont {D.}~\bibnamefont
  {Neuhauser}},\ }\href {https://aip.scitation.org/doi/10.1063/1.458900}
  {\bibfield  {journal} {\bibinfo  {journal} {J. Chem. Phys}\ }\textbf
  {\bibinfo {volume} {93}},\ \bibinfo {pages} {2611} (\bibinfo {year}
  {1990})}\BibitemShut {NoStop}%
\bibitem [{\citenamefont {Neuhauser}(1994)}]{neuhauser1994circumventing}%
  \BibitemOpen
  \bibfield  {author} {\bibinfo {author} {\bibfnamefont {D.}~\bibnamefont
  {Neuhauser}},\ }\href {https://aip.scitation.org/doi/10.1063/1.467224}
  {\bibfield  {journal} {\bibinfo  {journal} {J. Chem. Phys}\ }\textbf
  {\bibinfo {volume} {100}},\ \bibinfo {pages} {5076} (\bibinfo {year}
  {1994})}\BibitemShut {NoStop}%
\bibitem [{\citenamefont {McClean}\ \emph {et~al.}(2017)\citenamefont
  {McClean}, \citenamefont {Kimchi-Schwartz}, \citenamefont {Carter},\ and\
  \citenamefont {De~Jong}}]{mcclean2017hybrid}%
  \BibitemOpen
  \bibfield  {author} {\bibinfo {author} {\bibfnamefont {J.~R.}\ \bibnamefont
  {McClean}}, \bibinfo {author} {\bibfnamefont {M.~E.}\ \bibnamefont
  {Kimchi-Schwartz}}, \bibinfo {author} {\bibfnamefont {J.}~\bibnamefont
  {Carter}},\ and\ \bibinfo {author} {\bibfnamefont {W.~A.}\ \bibnamefont
  {De~Jong}},\ }\href
  {https://journals.aps.org/pra/abstract/10.1103/PhysRevA.95.042308} {\bibfield
   {journal} {\bibinfo  {journal} {Phys. Rev. A}\ }\textbf {\bibinfo {volume}
  {95}},\ \bibinfo {pages} {042308} (\bibinfo {year} {2017})}\BibitemShut
  {NoStop}%
\bibitem [{\citenamefont {Motta}\ \emph {et~al.}(2020)\citenamefont {Motta},
  \citenamefont {Sun}, \citenamefont {Tan}, \citenamefont {O’Rourke},
  \citenamefont {Ye}, \citenamefont {Minnich}, \citenamefont {Brandao},\ and\
  \citenamefont {Chan}}]{motta2020determining}%
  \BibitemOpen
  \bibfield  {author} {\bibinfo {author} {\bibfnamefont {M.}~\bibnamefont
  {Motta}}, \bibinfo {author} {\bibfnamefont {C.}~\bibnamefont {Sun}}, \bibinfo
  {author} {\bibfnamefont {A.~T.}\ \bibnamefont {Tan}}, \bibinfo {author}
  {\bibfnamefont {M.~J.}\ \bibnamefont {O’Rourke}}, \bibinfo {author}
  {\bibfnamefont {E.}~\bibnamefont {Ye}}, \bibinfo {author} {\bibfnamefont
  {A.~J.}\ \bibnamefont {Minnich}}, \bibinfo {author} {\bibfnamefont {F.~G.}\
  \bibnamefont {Brandao}},\ and\ \bibinfo {author} {\bibfnamefont {G.~K.-L.}\
  \bibnamefont {Chan}},\ }\href
  {https://www.nature.com/articles/s41567-019-0704-4} {\bibfield  {journal}
  {\bibinfo  {journal} {Nat. Phys}\ }\textbf {\bibinfo {volume} {16}},\
  \bibinfo {pages} {205} (\bibinfo {year} {2020})}\BibitemShut {NoStop}%
\bibitem [{\citenamefont {Ollitrault}\ \emph {et~al.}(2020)\citenamefont
  {Ollitrault}, \citenamefont {Kandala}, \citenamefont {Chen}, \citenamefont
  {Barkoutsos}, \citenamefont {Mezzacapo}, \citenamefont {Pistoia},
  \citenamefont {Sheldon}, \citenamefont {Woerner}, \citenamefont {Gambetta},\
  and\ \citenamefont {Tavernelli}}]{ollitrault2020quantum}%
  \BibitemOpen
  \bibfield  {author} {\bibinfo {author} {\bibfnamefont {P.~J.}\ \bibnamefont
  {Ollitrault}}, \bibinfo {author} {\bibfnamefont {A.}~\bibnamefont {Kandala}},
  \bibinfo {author} {\bibfnamefont {C.-F.}\ \bibnamefont {Chen}}, \bibinfo
  {author} {\bibfnamefont {P.~K.}\ \bibnamefont {Barkoutsos}}, \bibinfo
  {author} {\bibfnamefont {A.}~\bibnamefont {Mezzacapo}}, \bibinfo {author}
  {\bibfnamefont {M.}~\bibnamefont {Pistoia}}, \bibinfo {author} {\bibfnamefont
  {S.}~\bibnamefont {Sheldon}}, \bibinfo {author} {\bibfnamefont
  {S.}~\bibnamefont {Woerner}}, \bibinfo {author} {\bibfnamefont {J.~M.}\
  \bibnamefont {Gambetta}},\ and\ \bibinfo {author} {\bibfnamefont
  {I.}~\bibnamefont {Tavernelli}},\ }\href
  {https://journals.aps.org/prresearch/abstract/10.1103/PhysRevResearch.2.043140}
  {\bibfield  {journal} {\bibinfo  {journal} {Phys. Rev. Research}\ }\textbf
  {\bibinfo {volume} {2}},\ \bibinfo {pages} {043140} (\bibinfo {year}
  {2020})}\BibitemShut {NoStop}%
\bibitem [{\citenamefont {Poulin}\ \emph {et~al.}(2015)\citenamefont {Poulin},
  \citenamefont {Hastings}, \citenamefont {Wecker}, \citenamefont {Wiebe},
  \citenamefont {Doberty},\ and\ \citenamefont {Troyer}}]{poulin2014trotter}%
  \BibitemOpen
  \bibfield  {author} {\bibinfo {author} {\bibfnamefont {D.}~\bibnamefont
  {Poulin}}, \bibinfo {author} {\bibfnamefont {M.~B.}\ \bibnamefont
  {Hastings}}, \bibinfo {author} {\bibfnamefont {D.}~\bibnamefont {Wecker}},
  \bibinfo {author} {\bibfnamefont {N.}~\bibnamefont {Wiebe}}, \bibinfo
  {author} {\bibfnamefont {A.~C.}\ \bibnamefont {Doberty}},\ and\ \bibinfo
  {author} {\bibfnamefont {M.}~\bibnamefont {Troyer}},\ }\href
  {https://dl.acm.org/doi/10.5555/2871401.2871402} {\bibfield  {journal}
  {\bibinfo  {journal} {Quant. Info. Comput}\ }\textbf {\bibinfo {volume}
  {15}},\ \bibinfo {pages} {361–384} (\bibinfo {year} {2015})}\BibitemShut
  {NoStop}%
\bibitem [{\citenamefont {Peng}\ and\ \citenamefont
  {Kowalski}(2017)}]{peng2017highly}%
  \BibitemOpen
  \bibfield  {author} {\bibinfo {author} {\bibfnamefont {B.}~\bibnamefont
  {Peng}}\ and\ \bibinfo {author} {\bibfnamefont {K.}~\bibnamefont
  {Kowalski}},\ }\href {https://pubs.acs.org/doi/10.1021/acs.jctc.7b00605}
  {\bibfield  {journal} {\bibinfo  {journal} {J. Chem. Theory Comput}\ }\textbf
  {\bibinfo {volume} {13}},\ \bibinfo {pages} {4179} (\bibinfo {year}
  {2017})}\BibitemShut {NoStop}%
\bibitem [{\citenamefont {Motta}\ \emph {et~al.}(2018)\citenamefont {Motta},
  \citenamefont {Ye}, \citenamefont {McClean}, \citenamefont {Li},
  \citenamefont {Minnich}, \citenamefont {Babbush},\ and\ \citenamefont
  {Chan}}]{motta2018low}%
  \BibitemOpen
  \bibfield  {author} {\bibinfo {author} {\bibfnamefont {M.}~\bibnamefont
  {Motta}}, \bibinfo {author} {\bibfnamefont {E.}~\bibnamefont {Ye}}, \bibinfo
  {author} {\bibfnamefont {J.~R.}\ \bibnamefont {McClean}}, \bibinfo {author}
  {\bibfnamefont {Z.}~\bibnamefont {Li}}, \bibinfo {author} {\bibfnamefont
  {A.~J.}\ \bibnamefont {Minnich}}, \bibinfo {author} {\bibfnamefont
  {R.}~\bibnamefont {Babbush}},\ and\ \bibinfo {author} {\bibfnamefont {G.~K.}\
  \bibnamefont {Chan}},\ }\href {https://arxiv.org/abs/1808.02625} {\bibfield
  {journal} {\bibinfo  {journal} {arXiv:1808.02625}\ } (\bibinfo {year}
  {2018})}\BibitemShut {NoStop}%
\bibitem [{\citenamefont {Motta}\ \emph {et~al.}(2019)\citenamefont {Motta},
  \citenamefont {Shee}, \citenamefont {Zhang},\ and\ \citenamefont
  {Chan}}]{motta2019efficient}%
  \BibitemOpen
  \bibfield  {author} {\bibinfo {author} {\bibfnamefont {M.}~\bibnamefont
  {Motta}}, \bibinfo {author} {\bibfnamefont {J.}~\bibnamefont {Shee}},
  \bibinfo {author} {\bibfnamefont {S.}~\bibnamefont {Zhang}},\ and\ \bibinfo
  {author} {\bibfnamefont {G.~K.-L.}\ \bibnamefont {Chan}},\ }\href
  {https://pubs.acs.org/doi/abs/10.1021/acs.jctc.8b00996} {\bibfield  {journal}
  {\bibinfo  {journal} {J. Chem. Theory Comput}\ }\textbf {\bibinfo {volume}
  {15}},\ \bibinfo {pages} {3510} (\bibinfo {year} {2019})}\BibitemShut
  {NoStop}%
\bibitem [{\citenamefont {Kivlichan}\ \emph {et~al.}(2018)\citenamefont
  {Kivlichan}, \citenamefont {McClean}, \citenamefont {Wiebe}, \citenamefont
  {Gidney}, \citenamefont {Aspuru-Guzik}, \citenamefont {Chan},\ and\
  \citenamefont {Babbush}}]{kivlichan2018quantum}%
  \BibitemOpen
  \bibfield  {author} {\bibinfo {author} {\bibfnamefont {I.~D.}\ \bibnamefont
  {Kivlichan}}, \bibinfo {author} {\bibfnamefont {J.}~\bibnamefont {McClean}},
  \bibinfo {author} {\bibfnamefont {N.}~\bibnamefont {Wiebe}}, \bibinfo
  {author} {\bibfnamefont {C.}~\bibnamefont {Gidney}}, \bibinfo {author}
  {\bibfnamefont {A.}~\bibnamefont {Aspuru-Guzik}}, \bibinfo {author}
  {\bibfnamefont {G.~K.-L.}\ \bibnamefont {Chan}},\ and\ \bibinfo {author}
  {\bibfnamefont {R.}~\bibnamefont {Babbush}},\ }\href
  {https://journals.aps.org/prl/abstract/10.1103/PhysRevLett.120.110501}
  {\bibfield  {journal} {\bibinfo  {journal} {Phys. Rev. Lett}\ }\textbf
  {\bibinfo {volume} {120}},\ \bibinfo {pages} {110501} (\bibinfo {year}
  {2018})}\BibitemShut {NoStop}%
\bibitem [{\citenamefont {Berry}\ \emph {et~al.}(2019)\citenamefont {Berry},
  \citenamefont {Gidney}, \citenamefont {Motta}, \citenamefont {McClean},\ and\
  \citenamefont {Babbush}}]{berry2019qubitization}%
  \BibitemOpen
  \bibfield  {author} {\bibinfo {author} {\bibfnamefont {D.~W.}\ \bibnamefont
  {Berry}}, \bibinfo {author} {\bibfnamefont {C.}~\bibnamefont {Gidney}},
  \bibinfo {author} {\bibfnamefont {M.}~\bibnamefont {Motta}}, \bibinfo
  {author} {\bibfnamefont {J.~R.}\ \bibnamefont {McClean}},\ and\ \bibinfo
  {author} {\bibfnamefont {R.}~\bibnamefont {Babbush}},\ }\href
  {https://quantum-journal.org/papers/q-2019-12-02-208/} {\bibfield  {journal}
  {\bibinfo  {journal} {Quantum}\ }\textbf {\bibinfo {volume} {3}},\ \bibinfo
  {pages} {208} (\bibinfo {year} {2019})}\BibitemShut {NoStop}%
\bibitem [{\citenamefont {Matsuzawa}\ and\ \citenamefont
  {Kurashige}(2020)}]{matsuzawa2020jastrow}%
  \BibitemOpen
  \bibfield  {author} {\bibinfo {author} {\bibfnamefont {Y.}~\bibnamefont
  {Matsuzawa}}\ and\ \bibinfo {author} {\bibfnamefont {Y.}~\bibnamefont
  {Kurashige}},\ }\href
  {https://pubs.acs.org/doi/full/10.1021/acs.jctc.9b00963} {\bibfield
  {journal} {\bibinfo  {journal} {J. Chem. Theory Comput}\ }\textbf {\bibinfo
  {volume} {16}},\ \bibinfo {pages} {944} (\bibinfo {year} {2020})}\BibitemShut
  {NoStop}%
\bibitem [{\citenamefont {Huggins}\ \emph {et~al.}(2021)\citenamefont
  {Huggins}, \citenamefont {McClean}, \citenamefont {Rubin}, \citenamefont
  {Jiang}, \citenamefont {Wiebe}, \citenamefont {Whaley},\ and\ \citenamefont
  {Babbush}}]{huggins2021efficient}%
  \BibitemOpen
  \bibfield  {author} {\bibinfo {author} {\bibfnamefont {W.~J.}\ \bibnamefont
  {Huggins}}, \bibinfo {author} {\bibfnamefont {J.~R.}\ \bibnamefont
  {McClean}}, \bibinfo {author} {\bibfnamefont {N.~C.}\ \bibnamefont {Rubin}},
  \bibinfo {author} {\bibfnamefont {Z.}~\bibnamefont {Jiang}}, \bibinfo
  {author} {\bibfnamefont {N.}~\bibnamefont {Wiebe}}, \bibinfo {author}
  {\bibfnamefont {K.~B.}\ \bibnamefont {Whaley}},\ and\ \bibinfo {author}
  {\bibfnamefont {R.}~\bibnamefont {Babbush}},\ }\href
  {https://www.nature.com/articles/s41534-020-00341-7} {\bibfield  {journal}
  {\bibinfo  {journal} {npj Quantum Inf}\ }\textbf {\bibinfo {volume} {7}},\
  \bibinfo {pages} {1} (\bibinfo {year} {2021})}\BibitemShut {NoStop}%
\bibitem [{\citenamefont {Helgaker}\ \emph {et~al.}(2014)\citenamefont
  {Helgaker}, \citenamefont {Jorgensen},\ and\ \citenamefont
  {Olsen}}]{helgaker2014molecular}%
  \BibitemOpen
  \bibfield  {author} {\bibinfo {author} {\bibfnamefont {T.}~\bibnamefont
  {Helgaker}}, \bibinfo {author} {\bibfnamefont {P.}~\bibnamefont
  {Jorgensen}},\ and\ \bibinfo {author} {\bibfnamefont {J.}~\bibnamefont
  {Olsen}},\ }\href@noop {} {\emph {\bibinfo {title} {Molecular
  electronic-structure theory}}}\ (\bibinfo  {publisher} {John Wiley \& Sons},\
  \bibinfo {year} {2014})\BibitemShut {NoStop}%
\bibitem [{\citenamefont {Verteletskyi}\ \emph {et~al.}(2020)\citenamefont
  {Verteletskyi}, \citenamefont {Yen},\ and\ \citenamefont
  {Izmaylov}}]{verteletskyi2020measurement}%
  \BibitemOpen
  \bibfield  {author} {\bibinfo {author} {\bibfnamefont {V.}~\bibnamefont
  {Verteletskyi}}, \bibinfo {author} {\bibfnamefont {T.-C.}\ \bibnamefont
  {Yen}},\ and\ \bibinfo {author} {\bibfnamefont {A.~F.}\ \bibnamefont
  {Izmaylov}},\ }\href {https://aip.scitation.org/doi/10.1063/1.5141458}
  {\bibfield  {journal} {\bibinfo  {journal} {J. Chem. Phys}\ }\textbf
  {\bibinfo {volume} {152}},\ \bibinfo {pages} {124114} (\bibinfo {year}
  {2020})}\BibitemShut {NoStop}%
\bibitem [{\citenamefont {Gokhale}\ \emph {et~al.}(2019)\citenamefont
  {Gokhale}, \citenamefont {Angiuli}, \citenamefont {Ding}, \citenamefont
  {Gui}, \citenamefont {Tomesh}, \citenamefont {Suchara}, \citenamefont
  {Martonosi},\ and\ \citenamefont {Chong}}]{gokhale2019minimizing}%
  \BibitemOpen
  \bibfield  {author} {\bibinfo {author} {\bibfnamefont {P.}~\bibnamefont
  {Gokhale}}, \bibinfo {author} {\bibfnamefont {O.}~\bibnamefont {Angiuli}},
  \bibinfo {author} {\bibfnamefont {Y.}~\bibnamefont {Ding}}, \bibinfo {author}
  {\bibfnamefont {K.}~\bibnamefont {Gui}}, \bibinfo {author} {\bibfnamefont
  {T.}~\bibnamefont {Tomesh}}, \bibinfo {author} {\bibfnamefont
  {M.}~\bibnamefont {Suchara}}, \bibinfo {author} {\bibfnamefont
  {M.}~\bibnamefont {Martonosi}},\ and\ \bibinfo {author} {\bibfnamefont
  {F.~T.}\ \bibnamefont {Chong}},\ }\href {https://arxiv.org/abs/1907.13623}
  {\bibfield  {journal} {\bibinfo  {journal} {arXiv preprint arXiv:1907.13623}\
  } (\bibinfo {year} {2019})}\BibitemShut {NoStop}%
\bibitem [{Note1()}]{Note1}%
  \BibitemOpen
  \bibinfo {note} {Note that this notation of double factorization $\rightarrow
  $ DF presents an unfortunate elision with the popular and related density
  fitting representation, which is often denoted as DF in the literature. To
  avoid any issues, we will explicitly write out ``density fitting'' for the
  few times it is encountered in this manuscript.}\BibitemShut {Stop}%
\bibitem [{\citenamefont {Reck}\ \emph {et~al.}(1994)\citenamefont {Reck},
  \citenamefont {Zeilinger}, \citenamefont {Bernstein},\ and\ \citenamefont
  {Bertani}}]{reck1994experimental}%
  \BibitemOpen
  \bibfield  {author} {\bibinfo {author} {\bibfnamefont {M.}~\bibnamefont
  {Reck}}, \bibinfo {author} {\bibfnamefont {A.}~\bibnamefont {Zeilinger}},
  \bibinfo {author} {\bibfnamefont {H.~J.}\ \bibnamefont {Bernstein}},\ and\
  \bibinfo {author} {\bibfnamefont {P.}~\bibnamefont {Bertani}},\ }\href
  {https://journals.aps.org/prl/abstract/10.1103/PhysRevLett.73.58} {\bibfield
  {journal} {\bibinfo  {journal} {Phys. Rev. Lett}\ }\textbf {\bibinfo {volume}
  {73}},\ \bibinfo {pages} {58} (\bibinfo {year} {1994})}\BibitemShut {NoStop}%
\bibitem [{\citenamefont {Wecker}\ \emph {et~al.}(2015)\citenamefont {Wecker},
  \citenamefont {Hastings}, \citenamefont {Wiebe}, \citenamefont {Clark},
  \citenamefont {Nayak},\ and\ \citenamefont {Troyer}}]{wecker2015solving}%
  \BibitemOpen
  \bibfield  {author} {\bibinfo {author} {\bibfnamefont {D.}~\bibnamefont
  {Wecker}}, \bibinfo {author} {\bibfnamefont {M.~B.}\ \bibnamefont
  {Hastings}}, \bibinfo {author} {\bibfnamefont {N.}~\bibnamefont {Wiebe}},
  \bibinfo {author} {\bibfnamefont {B.~K.}\ \bibnamefont {Clark}}, \bibinfo
  {author} {\bibfnamefont {C.}~\bibnamefont {Nayak}},\ and\ \bibinfo {author}
  {\bibfnamefont {M.}~\bibnamefont {Troyer}},\ }\href
  {https://journals.aps.org/pra/abstract/10.1103/PhysRevA.92.062318} {\bibfield
   {journal} {\bibinfo  {journal} {Phys. Rev. A}\ }\textbf {\bibinfo {volume}
  {92}},\ \bibinfo {pages} {062318} (\bibinfo {year} {2015})}\BibitemShut
  {NoStop}%
\bibitem [{\citenamefont {Arute}\ \emph {et~al.}(2020)\citenamefont {Arute}
  \emph {et~al.}}]{google2020hartree}%
  \BibitemOpen
  \bibfield  {author} {\bibinfo {author} {\bibfnamefont {F.}~\bibnamefont
  {Arute}} \emph {et~al.} (\bibinfo {collaboration} {Google AI Quantum}),\
  }\href {https://science.sciencemag.org/content/369/6507/1084} {\bibfield
  {journal} {\bibinfo  {journal} {Science}\ }\textbf {\bibinfo {volume}
  {369}},\ \bibinfo {pages} {1084} (\bibinfo {year} {2020})}\BibitemShut
  {NoStop}%
\bibitem [{\citenamefont {Wilcox}(1967)}]{wilcox1967exponential}%
  \BibitemOpen
  \bibfield  {author} {\bibinfo {author} {\bibfnamefont {R.~M.}\ \bibnamefont
  {Wilcox}},\ }\href {https://aip.scitation.org/doi/10.1063/1.1705306}
  {\bibfield  {journal} {\bibinfo  {journal} {J. Math. Phys}\ }\textbf
  {\bibinfo {volume} {8}},\ \bibinfo {pages} {962} (\bibinfo {year}
  {1967})}\BibitemShut {NoStop}%
\bibitem [{\citenamefont {Parrish}\ \emph {et~al.}(2012)\citenamefont
  {Parrish}, \citenamefont {Hohenstein}, \citenamefont {Mart{\'\i}nez},\ and\
  \citenamefont {Sherrill}}]{parrish2012tensor}%
  \BibitemOpen
  \bibfield  {author} {\bibinfo {author} {\bibfnamefont {R.~M.}\ \bibnamefont
  {Parrish}}, \bibinfo {author} {\bibfnamefont {E.~G.}\ \bibnamefont
  {Hohenstein}}, \bibinfo {author} {\bibfnamefont {T.~J.}\ \bibnamefont
  {Mart{\'\i}nez}},\ and\ \bibinfo {author} {\bibfnamefont {C.~D.}\
  \bibnamefont {Sherrill}},\ }\href
  {https://aip.scitation.org/doi/10.1063/1.4768233} {\bibfield  {journal}
  {\bibinfo  {journal} {J. Chem. Phys}\ }\textbf {\bibinfo {volume} {137}},\
  \bibinfo {pages} {224106} (\bibinfo {year} {2012})}\BibitemShut {NoStop}%
\bibitem [{\citenamefont {Head-Gordon}\ and\ \citenamefont
  {Pople}(1988)}]{head1988optimization}%
  \BibitemOpen
  \bibfield  {author} {\bibinfo {author} {\bibfnamefont {M.}~\bibnamefont
  {Head-Gordon}}\ and\ \bibinfo {author} {\bibfnamefont {J.~A.}\ \bibnamefont
  {Pople}},\ }\href {https://pubs.acs.org/doi/10.1021/j100322a012} {\bibfield
  {journal} {\bibinfo  {journal} {J. Phys. Chem}\ }\textbf {\bibinfo {volume}
  {92}},\ \bibinfo {pages} {3063} (\bibinfo {year} {1988})}\BibitemShut
  {NoStop}%
\bibitem [{\citenamefont {Aharonov}\ \emph {et~al.}(2006)\citenamefont
  {Aharonov}, \citenamefont {Jones},\ and\ \citenamefont
  {Landau}}]{aharonov2009polynomial}%
  \BibitemOpen
  \bibfield  {author} {\bibinfo {author} {\bibfnamefont {D.}~\bibnamefont
  {Aharonov}}, \bibinfo {author} {\bibfnamefont {V.}~\bibnamefont {Jones}},\
  and\ \bibinfo {author} {\bibfnamefont {Z.}~\bibnamefont {Landau}},\ }in\
  \href {https://doi.org/10.1145/1132516.1132579} {\emph {\bibinfo {booktitle}
  {{Proc STOC}}}}\ (\bibinfo  {publisher} {ACM},\ \bibinfo {year} {2006})\ p.\
  \bibinfo {pages} {427–436}\BibitemShut {NoStop}%
\bibitem [{\citenamefont {Tran}\ \emph {et~al.}(2021)\citenamefont {Tran},
  \citenamefont {Su}, \citenamefont {Carney},\ and\ \citenamefont
  {Taylor}}]{tran2021faster}%
  \BibitemOpen
  \bibfield  {author} {\bibinfo {author} {\bibfnamefont {M.~C.}\ \bibnamefont
  {Tran}}, \bibinfo {author} {\bibfnamefont {Y.}~\bibnamefont {Su}}, \bibinfo
  {author} {\bibfnamefont {D.}~\bibnamefont {Carney}},\ and\ \bibinfo {author}
  {\bibfnamefont {J.~M.}\ \bibnamefont {Taylor}},\ }\href
  {https://dx.doi.org/10.1103/PRXQuantum.2.010323} {\bibfield  {journal}
  {\bibinfo  {journal} {Phys. Rev. X Quantum}\ }\textbf {\bibinfo {volume}
  {2}},\ \bibinfo {pages} {010323} (\bibinfo {year} {2021})}\BibitemShut
  {NoStop}%
\bibitem [{\citenamefont {Bonet-Monroig}\ \emph {et~al.}(2018)\citenamefont
  {Bonet-Monroig}, \citenamefont {Sagastizabal}, \citenamefont {Singh},\ and\
  \citenamefont {O'Brien}}]{bonet2018low}%
  \BibitemOpen
  \bibfield  {author} {\bibinfo {author} {\bibfnamefont {X.}~\bibnamefont
  {Bonet-Monroig}}, \bibinfo {author} {\bibfnamefont {R.}~\bibnamefont
  {Sagastizabal}}, \bibinfo {author} {\bibfnamefont {M.}~\bibnamefont
  {Singh}},\ and\ \bibinfo {author} {\bibfnamefont {T.}~\bibnamefont
  {O'Brien}},\ }\href
  {https://journals.aps.org/pra/abstract/10.1103/PhysRevA.98.062339} {\bibfield
   {journal} {\bibinfo  {journal} {Phys. Rev. A}\ }\textbf {\bibinfo {volume}
  {98}},\ \bibinfo {pages} {062339} (\bibinfo {year} {2018})}\BibitemShut
  {NoStop}%
\bibitem [{\citenamefont {McArdle}\ \emph {et~al.}(2019)\citenamefont
  {McArdle}, \citenamefont {Yuan},\ and\ \citenamefont
  {Benjamin}}]{mcardle2019error}%
  \BibitemOpen
  \bibfield  {author} {\bibinfo {author} {\bibfnamefont {S.}~\bibnamefont
  {McArdle}}, \bibinfo {author} {\bibfnamefont {X.}~\bibnamefont {Yuan}},\ and\
  \bibinfo {author} {\bibfnamefont {S.}~\bibnamefont {Benjamin}},\ }\href
  {https://journals.aps.org/prl/abstract/10.1103/PhysRevLett.122.180501}
  {\bibfield  {journal} {\bibinfo  {journal} {Phys. Rev. Lett}\ }\textbf
  {\bibinfo {volume} {122}},\ \bibinfo {pages} {180501} (\bibinfo {year}
  {2019})}\BibitemShut {NoStop}%
\bibitem [{\citenamefont {Aleksandrowicz}\ \emph {et~al.}(2019)\citenamefont
  {Aleksandrowicz}, \citenamefont {Alexander}, \citenamefont {Barkoutsos},
  \citenamefont {Bello}, \citenamefont {Ben-Haim}, \citenamefont {Bucher},
  \citenamefont {Cabrera-Hern{\'a}ndez}, \citenamefont {Carballo-Franquis},
  \citenamefont {Chen}, \citenamefont {Chen} \emph
  {et~al.}}]{aleksandrowicz2019qiskit}%
  \BibitemOpen
  \bibfield  {author} {\bibinfo {author} {\bibfnamefont {G.}~\bibnamefont
  {Aleksandrowicz}}, \bibinfo {author} {\bibfnamefont {T.}~\bibnamefont
  {Alexander}}, \bibinfo {author} {\bibfnamefont {P.}~\bibnamefont
  {Barkoutsos}}, \bibinfo {author} {\bibfnamefont {L.}~\bibnamefont {Bello}},
  \bibinfo {author} {\bibfnamefont {Y.}~\bibnamefont {Ben-Haim}}, \bibinfo
  {author} {\bibfnamefont {D.}~\bibnamefont {Bucher}}, \bibinfo {author}
  {\bibfnamefont {F.}~\bibnamefont {Cabrera-Hern{\'a}ndez}}, \bibinfo {author}
  {\bibfnamefont {J.}~\bibnamefont {Carballo-Franquis}}, \bibinfo {author}
  {\bibfnamefont {A.}~\bibnamefont {Chen}}, \bibinfo {author} {\bibfnamefont
  {C.}~\bibnamefont {Chen}}, \emph {et~al.},\ }\href
  {https://zenodo.org/record/2562111#.XhA8qi2ZPyI} {\bibfield  {journal}
  {\bibinfo  {journal} {Zenodo}\ } (\bibinfo {year} {2019})}\BibitemShut
  {NoStop}%
\bibitem [{\citenamefont {Cross}\ \emph {et~al.}(2019)\citenamefont {Cross},
  \citenamefont {Bishop}, \citenamefont {Sheldon}, \citenamefont {Nation},\
  and\ \citenamefont {Gambetta}}]{cross2019QV}%
  \BibitemOpen
  \bibfield  {author} {\bibinfo {author} {\bibfnamefont {A.~W.}\ \bibnamefont
  {Cross}}, \bibinfo {author} {\bibfnamefont {L.~S.}\ \bibnamefont {Bishop}},
  \bibinfo {author} {\bibfnamefont {S.}~\bibnamefont {Sheldon}}, \bibinfo
  {author} {\bibfnamefont {P.~D.}\ \bibnamefont {Nation}},\ and\ \bibinfo
  {author} {\bibfnamefont {J.~M.}\ \bibnamefont {Gambetta}},\ }\href
  {https://doi.org/10.1103/PhysRevA.100.032328} {\bibfield  {journal} {\bibinfo
   {journal} {Phys. Rev. A}\ }\textbf {\bibinfo {volume} {100}},\ \bibinfo
  {pages} {032328} (\bibinfo {year} {2019})}\BibitemShut {NoStop}%
\bibitem [{\citenamefont {\textit{ibmq$\_$guadalupe} v1.3.1}\ \emph
  {et~al.}(2020)\citenamefont {\textit{ibmq$\_$guadalupe} v1.3.1},
  \citenamefont {\textit{ibmq$\_$montreal} v1.3.1},\ and\ \citenamefont
  {\textit{ibmq$\_$mumbai} v1.3.3}}]{IBMQDevices}%
  \BibitemOpen
  \bibfield  {author} {\bibinfo {author} {\bibnamefont
  {\textit{ibmq$\_$guadalupe} v1.3.1}}, \bibinfo {author} {\bibnamefont
  {\textit{ibmq$\_$montreal} v1.3.1}},\ and\ \bibinfo {author} {\bibnamefont
  {\textit{ibmq$\_$mumbai} v1.3.3}},\ }\href@noop {} {}\bibinfo {howpublished}
  {IBM Quantum Team, Retrieved from https://quantum-computing.ibm.com}
  (\bibinfo {year} {2020})\BibitemShut {NoStop}%
\bibitem [{\citenamefont {Whitten}(1973)}]{Whitten:1973:4496}%
  \BibitemOpen
  \bibfield  {author} {\bibinfo {author} {\bibfnamefont {J.~L.}\ \bibnamefont
  {Whitten}},\ }\href {https://doi.org/NONE} {\bibfield  {journal} {\bibinfo
  {journal} {J. Chem. Phys.}\ }\textbf {\bibinfo {volume} {58}},\ \bibinfo
  {pages} {4496} (\bibinfo {year} {1973})}\BibitemShut {NoStop}%
\bibitem [{\citenamefont {Dunlap}\ \emph {et~al.}(1977)\citenamefont {Dunlap},
  \citenamefont {Connolly},\ and\ \citenamefont {Sabin}}]{Dunlap:1977:81}%
  \BibitemOpen
  \bibfield  {author} {\bibinfo {author} {\bibfnamefont {B.~I.}\ \bibnamefont
  {Dunlap}}, \bibinfo {author} {\bibfnamefont {J.~W.~D.}\ \bibnamefont
  {Connolly}},\ and\ \bibinfo {author} {\bibfnamefont {J.~R.}\ \bibnamefont
  {Sabin}},\ }\href {https://doi.org/Transition-metal Atoms - Nickel Atom and
  Nickel H} {\bibfield  {journal} {\bibinfo  {journal} {Int. J. Quantum Chem.
  Symp.}\ }\textbf {\bibinfo {volume} {12}},\ \bibinfo {pages} {81} (\bibinfo
  {year} {1977})}\BibitemShut {NoStop}%
\bibitem [{\citenamefont {Dunlap}\ \emph {et~al.}(1979)\citenamefont {Dunlap},
  \citenamefont {Connolly},\ and\ \citenamefont {Sabin}}]{Dunlap:1979:3396}%
  \BibitemOpen
  \bibfield  {author} {\bibinfo {author} {\bibfnamefont {B.~I.}\ \bibnamefont
  {Dunlap}}, \bibinfo {author} {\bibfnamefont {J.~W.~D.}\ \bibnamefont
  {Connolly}},\ and\ \bibinfo {author} {\bibfnamefont {J.~R.}\ \bibnamefont
  {Sabin}},\ }\href {https://doi.org/10.1063/1.438728} {\bibfield  {journal}
  {\bibinfo  {journal} {J. Chem. Phys.}\ }\textbf {\bibinfo {volume} {71}},\
  \bibinfo {pages} {3396} (\bibinfo {year} {1979})}\BibitemShut {NoStop}%
\bibitem [{\citenamefont {Feyereisen}\ \emph {et~al.}(1993)\citenamefont
  {Feyereisen}, \citenamefont {Fitzgerald},\ and\ \citenamefont
  {Komornicki}}]{Feyereisen:1993:359}%
  \BibitemOpen
  \bibfield  {author} {\bibinfo {author} {\bibfnamefont {M.}~\bibnamefont
  {Feyereisen}}, \bibinfo {author} {\bibfnamefont {G.}~\bibnamefont
  {Fitzgerald}},\ and\ \bibinfo {author} {\bibfnamefont {A.}~\bibnamefont
  {Komornicki}},\ }\href
  {https://www.sciencedirect.com/science/article/abs/pii/000926149387156W?via%3Dihub}
  {\bibfield  {journal} {\bibinfo  {journal} {Chem. Phys. Lett.}\ }\textbf
  {\bibinfo {volume} {208}},\ \bibinfo {pages} {359} (\bibinfo {year}
  {1993})}\BibitemShut {NoStop}%
\bibitem [{\citenamefont {Komornicki}\ and\ \citenamefont
  {Fitzgerald}(1993)}]{Komornicki:1993:1398}%
  \BibitemOpen
  \bibfield  {author} {\bibinfo {author} {\bibfnamefont {A.}~\bibnamefont
  {Komornicki}}\ and\ \bibinfo {author} {\bibfnamefont {G.}~\bibnamefont
  {Fitzgerald}},\ }\href {https://doi.org/http://dx.doi.org/10.1063/1.465054}
  {\bibfield  {journal} {\bibinfo  {journal} {J. Chem. Phys.}\ }\textbf
  {\bibinfo {volume} {98}},\ \bibinfo {pages} {1398} (\bibinfo {year}
  {1993})}\BibitemShut {NoStop}%
\bibitem [{\citenamefont {Vahtras}\ \emph {et~al.}(1993)\citenamefont
  {Vahtras}, \citenamefont {Alml{\"o}f},\ and\ \citenamefont
  {Feyereisen}}]{Vahtras:1993:514}%
  \BibitemOpen
  \bibfield  {author} {\bibinfo {author} {\bibfnamefont {O.}~\bibnamefont
  {Vahtras}}, \bibinfo {author} {\bibfnamefont {J.}~\bibnamefont
  {Alml{\"o}f}},\ and\ \bibinfo {author} {\bibfnamefont {M.~W.}\ \bibnamefont
  {Feyereisen}},\ }\href {https://aip.scitation.org/doi/10.1063/1.2956507}
  {\bibfield  {journal} {\bibinfo  {journal} {Chem. Phys. Lett.}\ }\textbf
  {\bibinfo {volume} {213}},\ \bibinfo {pages} {514} (\bibinfo {year}
  {1993})}\BibitemShut {NoStop}%
\bibitem [{\citenamefont {Rendell}\ and\ \citenamefont
  {Lee}(1994)}]{Rendell:1994:400}%
  \BibitemOpen
  \bibfield  {author} {\bibinfo {author} {\bibfnamefont {A.~P.}\ \bibnamefont
  {Rendell}}\ and\ \bibinfo {author} {\bibfnamefont {T.~J.}\ \bibnamefont
  {Lee}},\ }\href {https://doi.org/INTN} {\bibfield  {journal} {\bibinfo
  {journal} {J. Chem. Phys.}\ }\textbf {\bibinfo {volume} {101}},\ \bibinfo
  {pages} {400} (\bibinfo {year} {1994})}\BibitemShut {NoStop}%
\bibitem [{\citenamefont {Kendall}\ and\ \citenamefont
  {Fruchtl}(1997)}]{Kendall:1997:158}%
  \BibitemOpen
  \bibfield  {author} {\bibinfo {author} {\bibfnamefont {R.~A.}\ \bibnamefont
  {Kendall}}\ and\ \bibinfo {author} {\bibfnamefont {H.~A.}\ \bibnamefont
  {Fruchtl}},\ }\href {https://doi.org/NONE} {\bibfield  {journal} {\bibinfo
  {journal} {Theor. Chem. Acc.}\ }\textbf {\bibinfo {volume} {97}},\ \bibinfo
  {pages} {158} (\bibinfo {year} {1997})}\BibitemShut {NoStop}%
\bibitem [{\citenamefont {Weigend}(2002)}]{Weigend:2002:4285}%
  \BibitemOpen
  \bibfield  {author} {\bibinfo {author} {\bibfnamefont {F.}~\bibnamefont
  {Weigend}},\ }\href {https://doi.org/10.1039/B204199P} {\bibfield  {journal}
  {\bibinfo  {journal} {Phys. Chem. Chem. Phys.}\ }\textbf {\bibinfo {volume}
  {4}},\ \bibinfo {pages} {4285} (\bibinfo {year} {2002})}\BibitemShut
  {NoStop}%
\bibitem [{\citenamefont {Beebe}\ and\ \citenamefont
  {Linderberg}(1977)}]{Beebe:1977:683}%
  \BibitemOpen
  \bibfield  {author} {\bibinfo {author} {\bibfnamefont {N.~H.~F.}\
  \bibnamefont {Beebe}}\ and\ \bibinfo {author} {\bibfnamefont
  {J.}~\bibnamefont {Linderberg}},\ }\href {https://doi.org/Two-electron
  Integrals in Molecular Calculations} {\bibfield  {journal} {\bibinfo
  {journal} {Int. J. Quantum Chem.}\ }\textbf {\bibinfo {volume} {12}},\
  \bibinfo {pages} {683} (\bibinfo {year} {1977})}\BibitemShut {NoStop}%
\bibitem [{\citenamefont {Roeggen}\ and\ \citenamefont
  {Wisloff-Nilssen}(1986)}]{Roeggen:1986:154}%
  \BibitemOpen
  \bibfield  {author} {\bibinfo {author} {\bibfnamefont {I.}~\bibnamefont
  {Roeggen}}\ and\ \bibinfo {author} {\bibfnamefont {E.}~\bibnamefont
  {Wisloff-Nilssen}},\ }\href {https://doi.org/GEXT} {\bibfield  {journal}
  {\bibinfo  {journal} {Chem. Phys. Lett.}\ }\textbf {\bibinfo {volume}
  {132}},\ \bibinfo {pages} {154} (\bibinfo {year} {1986})}\BibitemShut
  {NoStop}%
\bibitem [{\citenamefont {Koch}\ \emph {et~al.}(2003)\citenamefont {Koch},
  \citenamefont {de~Meras},\ and\ \citenamefont {Pedersen}}]{Koch:2003:9481}%
  \BibitemOpen
  \bibfield  {author} {\bibinfo {author} {\bibfnamefont {H.}~\bibnamefont
  {Koch}}, \bibinfo {author} {\bibfnamefont {A.~S.}\ \bibnamefont {de~Meras}},\
  and\ \bibinfo {author} {\bibfnamefont {T.~B.}\ \bibnamefont {Pedersen}},\
  }\href {https://doi.org/10.1063/1.1578621} {\bibfield  {journal} {\bibinfo
  {journal} {J. Chem. Phys.}\ }\textbf {\bibinfo {volume} {118}},\ \bibinfo
  {pages} {9481} (\bibinfo {year} {2003})}\BibitemShut {NoStop}%
\bibitem [{\citenamefont {Aquilante}\ \emph {et~al.}(2007)\citenamefont
  {Aquilante}, \citenamefont {Pedersen},\ and\ \citenamefont
  {Lindh}}]{Aquilante:2007:194106}%
  \BibitemOpen
  \bibfield  {author} {\bibinfo {author} {\bibfnamefont {F.}~\bibnamefont
  {Aquilante}}, \bibinfo {author} {\bibfnamefont {T.~B.}\ \bibnamefont
  {Pedersen}},\ and\ \bibinfo {author} {\bibfnamefont {R.}~\bibnamefont
  {Lindh}},\ }\href {https://doi.org/10.1063/1.2736701} {\bibfield  {journal}
  {\bibinfo  {journal} {J. Chem. Phys.}\ }\textbf {\bibinfo {volume} {126}},\
  \bibinfo {pages} {194106} (\bibinfo {year} {2007})}\BibitemShut {NoStop}%
\bibitem [{\citenamefont {Aquilante}\ \emph {et~al.}(2009)\citenamefont
  {Aquilante}, \citenamefont {Gagliardi}, \citenamefont {Pedersen},\ and\
  \citenamefont {Lindh}}]{Aquilante:2009:154107}%
  \BibitemOpen
  \bibfield  {author} {\bibinfo {author} {\bibfnamefont {F.}~\bibnamefont
  {Aquilante}}, \bibinfo {author} {\bibfnamefont {L.}~\bibnamefont
  {Gagliardi}}, \bibinfo {author} {\bibfnamefont {T.~B.}\ \bibnamefont
  {Pedersen}},\ and\ \bibinfo {author} {\bibfnamefont {R.}~\bibnamefont
  {Lindh}},\ }\href {https://doi.org/10.1063/1.3116784} {\bibfield  {journal}
  {\bibinfo  {journal} {J. Chem. Phys.}\ }\textbf {\bibinfo {volume} {130}},\
  \bibinfo {pages} {154107} (\bibinfo {year} {2009})}\BibitemShut {NoStop}%
\bibitem [{\citenamefont {Hohenstein}\ \emph {et~al.}(2012)\citenamefont
  {Hohenstein}, \citenamefont {Parrish},\ and\ \citenamefont
  {Mart{\'\i}nez}}]{hohenstein2012tensor}%
  \BibitemOpen
  \bibfield  {author} {\bibinfo {author} {\bibfnamefont {E.~G.}\ \bibnamefont
  {Hohenstein}}, \bibinfo {author} {\bibfnamefont {R.~M.}\ \bibnamefont
  {Parrish}},\ and\ \bibinfo {author} {\bibfnamefont {T.~J.}\ \bibnamefont
  {Mart{\'\i}nez}},\ }\href
  {https://aip.scitation.org/doi/abs/10.1063/1.4732310} {\bibfield  {journal}
  {\bibinfo  {journal} {J. Chem. Phys}\ }\textbf {\bibinfo {volume} {137}},\
  \bibinfo {pages} {044103} (\bibinfo {year} {2012})}\BibitemShut {NoStop}%
\bibitem [{\citenamefont {Parrish}\ \emph {et~al.}(2013)\citenamefont
  {Parrish}, \citenamefont {Hohenstein}, \citenamefont {Schunck}, \citenamefont
  {Sherrill},\ and\ \citenamefont {Mart{\'\i}nez}}]{parrish2013exact}%
  \BibitemOpen
  \bibfield  {author} {\bibinfo {author} {\bibfnamefont {R.~M.}\ \bibnamefont
  {Parrish}}, \bibinfo {author} {\bibfnamefont {E.~G.}\ \bibnamefont
  {Hohenstein}}, \bibinfo {author} {\bibfnamefont {N.~F.}\ \bibnamefont
  {Schunck}}, \bibinfo {author} {\bibfnamefont {C.~D.}\ \bibnamefont
  {Sherrill}},\ and\ \bibinfo {author} {\bibfnamefont {T.~J.}\ \bibnamefont
  {Mart{\'\i}nez}},\ }\href
  {https://journals.aps.org/prl/abstract/10.1103/PhysRevLett.111.132505}
  {\bibfield  {journal} {\bibinfo  {journal} {Phys. Rev. Lett}\ }\textbf
  {\bibinfo {volume} {111}},\ \bibinfo {pages} {132505} (\bibinfo {year}
  {2013})}\BibitemShut {NoStop}%
\bibitem [{\citenamefont {Lee}\ \emph {et~al.}(2020)\citenamefont {Lee},
  \citenamefont {Berry}, \citenamefont {Gidney}, \citenamefont {Huggins},
  \citenamefont {McClean}, \citenamefont {Wiebe},\ and\ \citenamefont
  {Babbush}}]{lee2020even}%
  \BibitemOpen
  \bibfield  {author} {\bibinfo {author} {\bibfnamefont {J.}~\bibnamefont
  {Lee}}, \bibinfo {author} {\bibfnamefont {D.}~\bibnamefont {Berry}}, \bibinfo
  {author} {\bibfnamefont {C.}~\bibnamefont {Gidney}}, \bibinfo {author}
  {\bibfnamefont {W.~J.}\ \bibnamefont {Huggins}}, \bibinfo {author}
  {\bibfnamefont {J.~R.}\ \bibnamefont {McClean}}, \bibinfo {author}
  {\bibfnamefont {N.}~\bibnamefont {Wiebe}},\ and\ \bibinfo {author}
  {\bibfnamefont {R.}~\bibnamefont {Babbush}},\ }\href
  {https://arxiv.org/abs/2011.03494} {\bibfield  {journal} {\bibinfo  {journal}
  {arXiv:2011.03494}\ } (\bibinfo {year} {2020})}\BibitemShut {NoStop}%
\bibitem [{\citenamefont {Bravyi}\ and\ \citenamefont
  {Kitaev}(2002)}]{bravyi2002fermionic}%
  \BibitemOpen
  \bibfield  {author} {\bibinfo {author} {\bibfnamefont {S.~B.}\ \bibnamefont
  {Bravyi}}\ and\ \bibinfo {author} {\bibfnamefont {A.~Y.}\ \bibnamefont
  {Kitaev}},\ }\href
  {https://www.sciencedirect.com/science/article/abs/pii/S0003491602962548}
  {\bibfield  {journal} {\bibinfo  {journal} {Ann. Phys}\ }\textbf {\bibinfo
  {volume} {298}},\ \bibinfo {pages} {210} (\bibinfo {year}
  {2002})}\BibitemShut {NoStop}%
\end{thebibliography}%

%\pagebreak
%\clearpage

\end{document}